\pdfoutput=1
\documentclass[prd,twocolumn,showpacs,amsmath,amssymb,floatfix,superscriptaddress]{revtex4-1}

\usepackage{graphicx}
\usepackage{dcolumn}
\usepackage{bm}
\usepackage{hyperref}

\usepackage{graphicx} 

\usepackage{graphicx}
\usepackage{bm}
\usepackage{suffix}
\usepackage{mathtools}

\DeclarePairedDelimiterX\MeijerM[3]{\lparen}{\rparen}%
{\begin{smallmatrix}#1 \\ #2\end{smallmatrix}\delimsize\vert\,#3}

\newcommand\MeijerG[8][]{%
  G^{\,#2,#3}_{#4,#5}\MeijerM[#1]{#6}{#7}{#8}}

\WithSuffix\newcommand\MeijerG*[7]{%
  G^{\,#1,#2}_{#3,#4}\MeijerM*{#5}{#6}{#7}}

\begin{document}

\title{Fast Summation of  Divergent Series and Resurgent Transseries in Quantum Field Theories from Meijer-G Approximants}
\author{H\'ector Mera}
\email{hypergeometric2f1@gmail.com}
\affiliation{Department of Physics and Astronomy, University of Delaware, Newark, DE 19716-2570, USA}
\affiliation{Department of Physics and Nanotechnology, Aalborg University, DK-9220 Aalborg {\O}st, Denmark }
\affiliation{Center for Nanostructured Graphene (CNG), DK-9220 Aalborg {\O}st, Denmark}
\author{Thomas G. Pedersen}
\affiliation{Department of Physics and Nanotechnology, Aalborg University, DK-9220 Aalborg {\O}st, Denmark }
\affiliation{Center for Nanostructured Graphene (CNG), DK-9220 Aalborg {\O}st, Denmark}
\author{Branislav K. Nikoli\'{c}}
\affiliation{Department of Physics and Astronomy, University of Delaware, Newark, DE 19716-2570, USA}
\date{\today} 

\begin{abstract}
We demonstrate that a  Meijer-G-function-based resummation approach  can be successfully applied to approximate the Borel sum of divergent series, and thus to approximate the Borel-\'Ecalle summation of  resurgent transseries in quantum field theory (QFT). The proposed method is shown to vastly outperform the conventional Borel-Pad\'e and Borel-Pad\'e-\'Ecalle summation methods. The resulting Meijer-G approximants are easily parameterized by means of a hypergeometric ansatz and can be thought of as a generalization to arbitrary order of the Borel-Hypergeometric method [Mera {\it et al.} Phys. Rev. Lett. {\bf 115}, 143001 (2015)]. Here we illustrate the ability of this technique in various examples from QFT, traditionally employed as benchmark models for resummation, such as: 0-dimensional $\phi^4$ theory, $\phi^4$ with degenerate minima, self-interacting QFT in 0-dimensions, and the computation of one- and two-instanton contributions in the quantum-mechanical double-well problem. 
\end{abstract}

\maketitle

\section{Introduction}\label{Sec:intro}
Perturbative expansions in quantum mechanics, quantum field theory (QFT) and string field theory often have {\em zero radius of convergence}, i.e., they are asymptotic\cite{Caliceti2007, Marino2014}. Optimal truncation of such series to a small number of terms can provide experimentally relevant results but only at sufficiently small coupling constant. This is exemplified by the asymptotic series for Stark effect of atoms and molecules in an external electric field where first and second order terms match measurements well, but only for very weak electric fields \cite{Bischel2007}, or by high precision calculations of multiloop Feynman diagrams in quantum electrodynamics \cite{Aoyama2015} --where the fine-structure constant is small. On the other hand, extracting physically relevant information from asymptotic series at larger coupling constants calls, almost invariably,  for resummation techniques as exemplified by the Stark effect~\cite{Mera2015,Pedersen2015} and field assisted excitonic ionization in novel materials~\cite{Pedersen2016},  anharmonic oscillators in quantum mechanics~\cite{Jentschura2009,Benders}, $\phi^4$ theory in  QFT  \cite{Kompaniets2017}, quantum chromodynamics \cite{Caprini2003}, 
string perturbation theory \cite{Gross1988, Grassi2015}, and diagrammatic Monte Carlo in condensed matter physics \cite{Prokofev2008}.  Given the ubiquity of divergent series in physics, research on summation techniques remains an active research area~\cite{Garcia2017, Costin2017}.

Conventional resummation is, however, not sufficient in the presence of the so-called Stokes phenomenon where different asymptotic expansions hold in different regions of the plane made up of  complexified expansion parameter values~\cite{Dingle1973, Berry1991, Aniceto2015a, Aniceto2015b, Cherman2015a, Marino2015, Santamaria2017}. Thus the Stokes phenomenon requires generally distinct resummations in each of these regions. This complexity is captured by  resurgent transseries \cite{Cherman2015a, Dorigoni2014,Heller2015,  Dunne2015,JentschuraMultinstanton}, which include both analytic polynomial terms and nonanalytic exponential and logarithmic terms. In principle, resurgent transseries offer a non-perturbative framework to reconstruct the original function, which has led to recent vigorous efforts to examine their promise in physically relevant examples, where different sectors are generated by non-perturbative semiclassical effects like instantons\cite{Marino2015, Santamaria2017, JentschuraMultinstanton}. 

Nonetheless, the resurgent transseries also need to be resummed in order to obtain a sensible result. However, the conventional Borel-Pad\'{e}-\'{E}calle  resummation used for this purpose typically requires a large number of terms~\cite{Santamaria2017, Cherman2015} from each sector in order to obtain reasonably accurate results  beyond the weak coupling regime. This makes it useless for problems in QFT~\cite{Aoyama2015, Kompaniets2017} or many-body perturbation theory in condensed matter physics~\cite{Prokofev2008} where $\lesssim 10$ orders are  available at best. Here we introduce a new algorithm which replaces the standard Pad\'e approximants in the Borel plane by more general and flexible hypergeometric functions (of, in principle, arbitrarily high order), thereby achieving great convergence acceleration towards the exact sum of a resurgent transseries. Hence, in this work, we replace the conventional  Borel-Pad\'e-\'Ecalle summation by  Borel-Hypergeometric-\'Ecalle summation whose approximants admit a representation in terms of Meijer-G functions which are easily parameterized. 

Prior to going into technical details, we highlight the power of our algorithm by noting that our approximants converge to the exact non-perturbative, ambiguity-free, partition function for 0-dimensional self-interacting QFT with just five orders, whereas Borel-Pad\'{e}-\'{E}calle resummation in Ref.~\onlinecite{Cherman2015}  needed tens of terms  to find a good approximation at intermediate coupling strength.  In practice only a handful of expansion coefficients are typically available: for instance the $\epsilon$-expansion for the $O(n)$-symmetric $\phi^4$ theory is only known to six-loop order~\cite{Kompaniets2017}, while the five-loop QCD beta function and anomalous dimension have been calculated only very recently~\cite{QCDbeta}.
It is then clear that, for a summation technique to be practical, it needs to be able to return accurate estimates of the sum of a divergent series with only a few coefficients. The high-accuracy at low orders of the Meijer-G approximants introduced in this work makes them suitable for practical applications. 

The paper is organized as follows. In Sec.~\ref{Sec:II} we introduce the algorithm to transform the  expansion coefficients  of a divergent expansion into a table of Meijer-G approximants. In Section~\ref{Sec:III} we apply our algorithm to sum four well-known examples in QFT: 0-dimensional $\phi^4$ theory, $\phi^4$ theory with degenerate minima, self-interacting 0-dimensional QFT and finally the computation of the one- and two-instanton contributions in the double-well problem of quantum mechanics.  While in the first of these examples we deal with a Borel-summable divergent expansion, the other three cases offer an opportunity to demonstrate the efficiency of our approach for the summation of transseries/multi-instanton expansions, i.e., those cases where the perturbation expansion is not Borel-summable. The Meijer-G summation method is shown to work well in all of these cases, providing a fast way to evaluating the Borel sum of a divergent series and massively outperforming the Borel-Pad\'e and Borel-Pad\'e-\'Ecalle approach. In the first three examples the Meijer-G approximants converge to the exact result at five-loop order, while in the latter case the convergence is slower---although fast when compared to Borel-Pad\'e-\'Ecalle approaches. Finally in Sec.~\ref{Sec:IV} we discuss the advantages and disadvantages of our approach as an alternative to traditional Borel-Pad\'e techniques. We end with the conclusions in Sec.~\ref{Sec:V}.

\section{Meijer-G Approximants}\label{Sec:II}
An efficient resummation technique should be capable of taking a handful of coefficients and return an accurate estimate of the sum of a divergent perturbation series. In this Section such a technique is introduced, providing an algorithm that efficiently transforms the coefficients of a divergent perturbation expansion into a table of Meijer-G functions which serve as approximants. For completeness we first briefly review the traditional Borel-Pad\'e resummation, emphasising the inherent difficulties faced by such method when it comes to summation ``on the cut''. We will then briefly review alternatives to Pad\'e and Borel-Pad\'e that make use of analytic continuation functions with a built-in branch cut, in particular the hypergeometric resummation method we introduced in Refs.~\onlinecite{Mera2015,Pedersen2015,Pedersen2016}. After reviewing these other approaches we finally introduce the algorithm to calculate Meijer-G approximants and highlight various properties that make this method extremely well-suited to yield inexpensive---and yet accurate---low order approximations to the sum of a divergent series.
\subsection{Borel-Pad\'e Resummation}
When it comes to sum divergent series, Borel-Pad\'e has become the dominant approach~\cite{Caliceti2007, Garcia2017,KleinertBook}. There are various reasons for the popularity of this approach but it can be argued that the most important of these is its algorithmic simplicity. The Borel-Pad\'e approach is in essence a simple recipe to transform the coefficients of a divergent expansion into a table of approximants, which approximate the Borel sum of a divergent series and are typically evaluated by numerical contour integration. Another advantage of this approach is that it relies on widely studied approaches: conditions for Borel-summability are by now well understood~\cite{KleinertBook, Graffi1970, Costin2008, Graffi1978} and the properties of Pad\'e approximants---used at a crucial step in the algorithm---are also very well known as they have been studied in depth for decades~\cite{Baker1996}. Given a divergent expansion $Z(g)\sim\sum_{n=0}^\infty z_n g^n$, where $z_n$ are the expansion coefficients and $g$ is the expansion parameter (``the coupling''), the algorithm to calculate the Borel sum of a divergent series  can be summarized as follows:

\underline{ \emph{Step 1: Borel transform.}}  Calculate the Borel-transformed coefficients: $b_n=z_n/n!$.

\underline{ \emph{Step 2: Summation in the Borel plane.}} Sum the series $B(\tau)\sim \sum_{n=0}^\infty b_n \tau^n$; this series is called the Borel-transformed series. The complex-$\tau$ plane is known as the Borel plane.

\underline{ \emph{Step 3: Laplace transform.}} The Borel sum of the series, $Z_B(g)$ is given by the Laplace transform 
\begin{equation}
Z_B(g)=\int_0^\infty e^{-\tau}B(\tau g) d\tau.
\end{equation}
The rationale behind the Borel summation method is  simple: the coefficients of a series with zero radius of convergence typically grow factorially at large orders. In the first step such factorial growth is removed and the Borel-transformed series is more tractable since it has a finite non-zero radius of convergence.  By summing the Borel-transformed series one finds the function $B(\tau)$ and then the Laplace transform can be calculated numerically to find the Borel sum $Z_B(g)$. The Borel-Pad\'e summation method is a practical algorithm to find, in principle, increasingly accurate approximations to the Borel sum. In this approach the second step above is specialized to Pad\'e summation:
\underline{ \emph{Step 2:  Pad\'e summation in the Borel plane.}} Use Pad\'e approximants to approximately  sum the series $B(\tau)\sim \sum_{n=0}^\infty b_n \tau^n$, from the knowledge of the partial sums to order $N$, i.e., $\sum_{n=0}^N b_n \tau^n$. In Pad\'e summation one approximates $B(\tau)$ by a rational function of $\tau$
\begin{equation}
B_{L/M}(\tau)=\frac{\sum_{n=0}^L p_n \tau^n}{1+\sum_{n=1}^M q_n\tau^n}.
\end{equation}
where $L+M=N$. Here the coefficients $p_n$ and $q_n$ are found by equating order by order the Taylor series of $B_{L/M}(\tau)$ around $\tau=0$ with the asymptotic expansion of $B(\tau)$, up to the desired order. These approximants are then used in step 3 of the algorithm to evaluate the Laplace transform and therefore to  find the $L/M$-Borel-Pad\'e approximant to the Borel sum of $Z(g)$, $Z_{B,L/M}(g)$, as
\begin{equation}
Z_{B,L/M}(g)=\int_0^\infty e^{-\tau}B_{L/M}(\tau g) d\tau.
\end{equation}
Clearly the Borel-Pad\'e method returns a table of approximants:  for instance by knowing the partial sums to second order ($N=2$) one can calculate $B_{1/0}$, $B_{0/1}$, $B_{2/0}$, $B_{1/1}$, and $B_{0/2}$.  

\begin{figure}
\center
\includegraphics[width=0.49\textwidth]{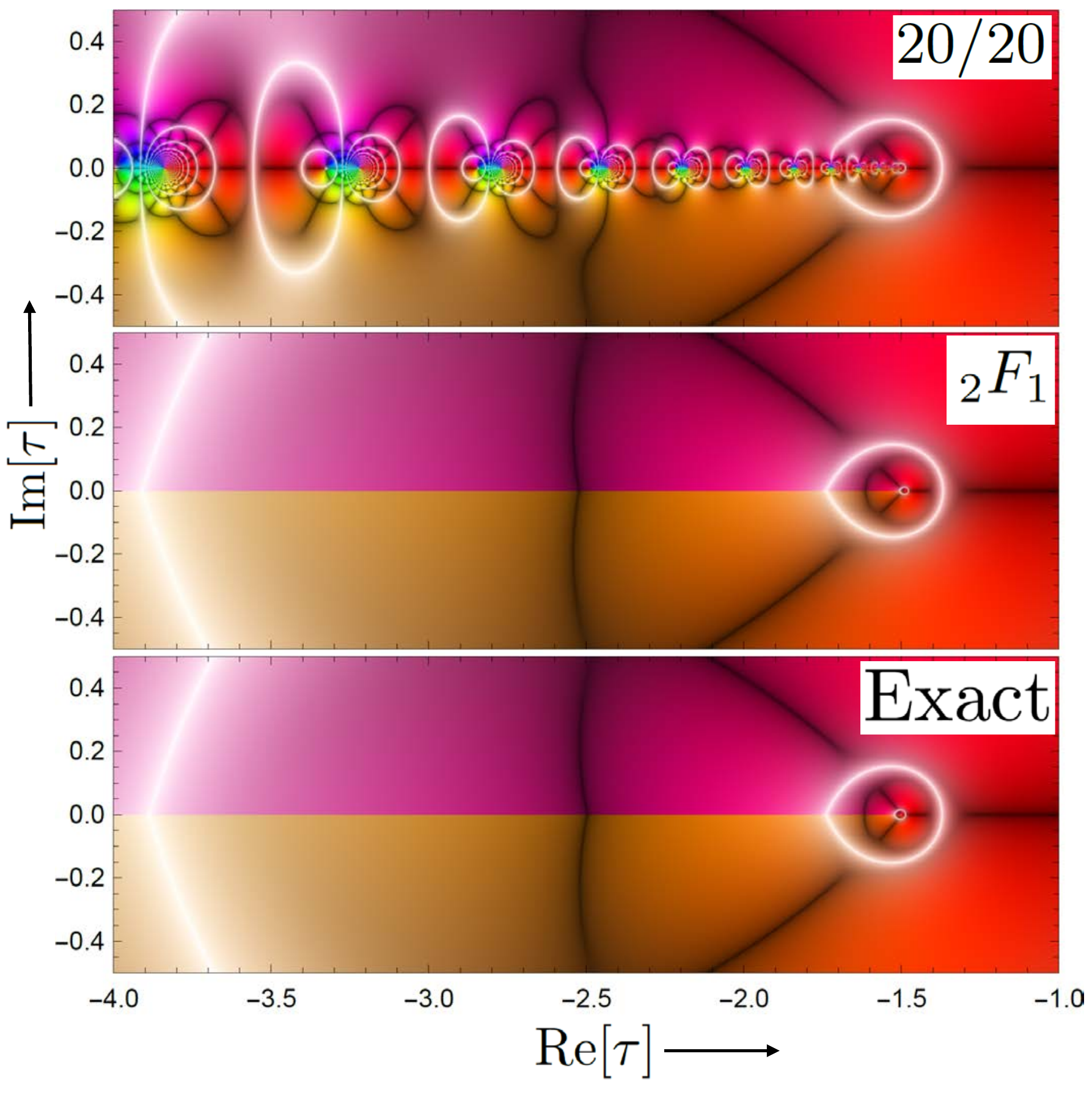}
\caption{Domain coloring plot of approximants in the Borel plane calculated for the partition function of $\phi^4$ theory in zero dimensions for $g=1$. Branch cuts are represented by a discontinuous change of color. Around poles one sees gradual colour changes, intersecting black lines  and  concentric white lines.  Top panel: $20/20$ Pad\'e approximant. Middle panel: hypergeometric $_2F_1$ approximant. Bottom panel: Exact. The Pad\'e approximant mimics the exact branch cut by placing poles next to each other and high orders are needed to accurately model the cut. In contrast, the third order $_2F_1$ approximant with a built-in branch cut very nearly reproduces the exact branch cut. }
\label{Fig:fig1}
\end{figure}

There are, however, various types of problems  for which Borel-Pad\'e approximants are not well suited. Pad\'e approximants are rational functions whose built-in singularities are poles. It is well-known that in  many situations the convergence of the Borel-transformed series is limited by the branch point singularity closest to the origin, the terminal point of a dense line of poles known as a branch cut. In such scenarios Borel-Pad\'e approximants can converge very slowly since many poles may be needed to properly mimic a branch cut. Thus, to overcome this difficulty with Borel-Pad\'e one has to replace Pad\'e approximants, adopting instead as approximants  functions that are able to mimic branch cuts in the Borel plane. The advantages of that strategy are highlighted in Fig.~\ref{Fig:fig1} where  we show a domain coloring plot of a function $B(\tau)$ in the Borel plane (corresponding to the partition function of 0-dimensional $\phi^4$ for $g=1$). We show the 20/20 Pad\'e approximant, $B_{20/20}(\tau)$ (top panel)  together with the exact $B(\tau)$ (bottom panel). We also show a hypergeometric $_2F_1$ approximant (middle panel) in the Borel plane which corresponds to the third order Meijer-G approximant, introduced in Sec.~II C. The Pad\'e approximant in the Borel plane attempts to reproduce the branch cut by brute force---placing poles next to each other along the negative $\tau$ axis. The Meijer-G approximant has a built-in branch cut in the Borel plane and it can thus very accurately mimic the branch cut using only three orders of perturbation theory. One has to carefully look at the  the details in Fig.~\ref{Fig:fig1} to see the very minor differences between the exact $B(\tau)$ and its third-order hypergeometric approximant.

\subsection{Hypergeometric Resummation}

Recently we have introduced hypergeometric resummation---a technique that enables summation ``on the cut'' using only a small number of expansion coefficients~\cite{Mera2015, Pedersen2015, Pedersen2016, Mera2016, Sanders2017}.   Various flavours  of this technique were applied to a variety of problems with good results: in particular it was shown how one could use low order data to derive accurate approximations to the decay rate in Stark-type problems~\cite{Mera2015, Pedersen2015, Pedersen2016}.  Typically the idea is to use hypergeometric functions to analytically continue divergent series. Originally the technique was devised by noticing the shortcomings of Pad\'e approximants when applied to problems in nonequilibrium nanoelectronics modelling  within the nonequilibrium Green's function theory~\cite{Mera2012, Cavassilas2013, Mera2013, Bescond2013}. As pointed out in Sec.~II~A one seeks to substitute Pad\'e approximants by more general functions that are equipped with branch cuts and can work well in those cases where Pad\'e approximants do not work or converge very slowly. Hypergeometric functions looked particularly promising as they are endowed with a branch cut and generalize geometric series (which can be summed exactly by Pad\'e approximants).  For instance given a divergent function $Z(g)\sim\sum_{n=0}^\infty z_n g^n$, with normalized zeroth order coefficient $z_0=1$, one can attempt to find a hypergeometric function\cite{dlmf, hyp2f1book}, 
$_2F_1(h_1,h_2,h_3,h_4 g)$, such that
\begin{equation}
Z(g)=\,_2F_1(h_1,h_2,h_3,h_4 g)+O(g^5).
\label{Eq:hyp2f1}
\end{equation}
Similar ideas were already considered  in Refs.~\onlinecite{KleinertBook, Kleinert1997,Jasch2001, Janke1995}, where the authors used it as part of an algorithm to find convergent strong coupling expansions (from divergent weak coupling ones). Similarly, in Refs.~\onlinecite{Yukalov1991,Yukalov1999}, the authors used products of hypergeometric functions as approximants. Foreshadowing all of these works are contributions by Stillinger and coworkers~\cite{Stillinger1979, Stillinger1983}, where early Borel-Hypergeometric approximants (typically $_1F_0$ hypergeometric functions in the Borel plane) are considered, as well as early versions of self-similar factor approximants and exponential-Borel approximants. Of the works mentioned, the work of Stillinger is the closest in philosophy to our own work, as he also considers the ratio test of series convergence as a starting point --as we do in Sec.~III C and we did in earlier work~\cite{Mera2016}.

The hypergeometric approximants we introduced in Refs.~\onlinecite{Mera2015, Pedersen2015,Pedersen2016, Mera2016} have a number of clear limitations, {\it which we overcome in this work} and which we enumerate below:
\begin{enumerate}
\item {\it Hypergeometric resummation is uncontrolled:} Hypergeometric $_2F_1$ approximants of third, fourth and fifth order can be constructed in various ways, but in Refs.~\onlinecite{Mera2015, Pedersen2015,Pedersen2016, Mera2016}  we did not give a recipe for  constructing a table of hypergeometric approximants. How does one parameterize and build, say, a 21st order hypergeometric approximant? In order to have control over any approximations one develops, it is fundamental to be able to increase the order of the approximation and to study the convergence properties of the approximants. \emph{In this work we provide a set of approximants that can, in principle, be computed at any order}.
\item {\it Difficult parameterization at large orders:} A natural attempt to provide a generalization to arbitrary orders would be to state that general hypergeometric  functions $_qF_p$~\cite{hyp2f1book, dlmf} constitute the approximant space. These functions contain $p+q+1$ parameters, $_pF_q(h_1,\ldots,h_p;h_{p+1},\ldots,h_{p+q}, h_{p+q+1}\tau)$, that need to be calculated by equating each order of the asymptotic expansion that one seeks to sum with the corresponding order of the expansion of the hypergeometric approximant. However one faces a degeneracy problem as all the hypergeometric functions obtained by permuting elements of each of the parameter sets $(h_1,\ldots,h_p)$ and $(h_{p+1},\ldots,h_{p+q})$ are one and the same. Hence when determining the parameters $h_i$ the computational time grows factorially with order --there is a factorially large number of solutions, all of which correspond to the same hypergeometric function. So calculations at the six-loop order are already very expensive and nearly impossible at the 7th loop order. The same problem is found with other approximants; for instance, self-similar factor approximants~\cite{Yukalov1991,Yukalov1999} have rarely been computed beyond the sixth loop order. \emph{In this work we put forward an algorithm that enables fast parameterization of hypergeometric approximants in the Borel plane, at arbitrarily high orders} .
\item {\it Inaccuracies for expansions with zero radius of convergence:}
It was noted  that for very small couplings the hypergeometric approximants (as described in Ref.~\onlinecite{Mera2015} ) gave inaccurate results. While these inaccuracies were exponentially small, they were conceptually important. The reason for them was that the radius of convergence of hypergeometric $_{p+1}F_p$ functions is not zero, while we were applying the hypergeometric approximants to problems with zero radius of convergence. For instance an approximant given by Eq.~\eqref{Eq:hyp2f1} above has radius of convergence $g_c=1/h_4$. When applied to series with zero radius of convergence the value of $h_4$ was typically found to be  very large but finite. Therefore the hypergeometric approximants had a Taylor series with a tiny, but non-zero, radius of convergence.  Hence, in Refs. \onlinecite{Pedersen2015, Pedersen2016} we used a different parameterization of the approximants,  which positioned the tip of the hypergeometric branch cut exactly at the origin and therefore alleviated this problem. However there we did not come up with a clear approach to compute similar approximants of higher order. \emph{The approximants derived in this work can have zero radius of convergence, therefore alleviating this difficulty found in our previous approach.}  
\end{enumerate}

Clearly there is a wide variety of problems for which Borel-Pad\'e resummation can be improved. However attempts to improve it can easily fall into severall pitfalls. In the case of hypergeometric resummation these were difficulties in both extending the approach to arbitrarily high orders and dealing with series with zero radius of convergence. These difficulties are largely surpassed by the Meijer-G approximants we introduce next.

\subsection{Meijer-G Resummation}
We now present an  algorithm to transform the low order coefficients of a divergent perturbation expansion, $Z(g)\sim \sum_{n=0}^\infty z_n g^n$ (with normalized coefficients, $z_0=1$), into a table of approximants to its Borel sum. The algorithm  consists of four easy steps.  For ease of presentation, in this work we will compute mostly \emph{odd-order} approximants, giving a short description of the algorithm to compute even-order approximants below.

\underline{ \emph{Step 1: Borel transform.}} Imagine that we know only $N$ coefficients, $z_0,z_1,\ldots,z_N$. In this step we compute the Borel-transformed coefficients $b_n=z_n/n!$ together $N$ ratios of consecutive Borel-transformed coefficients, $r(n)=b_{n+1}/b_n$. Here we will assume that $N$ is an odd number.

\underline{\emph{Step 2: Hypergeometric ansatz.}}  We make the ansatz that $r(n)$  is a rational function of $n$. Thus we define a rational function of $n$, $r_N(n)$ as
\begin{equation}
r_N(n)=\frac{\sum_{m=0}^l p_m n^m}{1+\sum_{m=1}^l q_m n^m},
\end{equation}
where $l=(N-1)/2$ and the $N$ unknown parameters $p_m$ and $q_m$ are \emph{uniquely} determined by the $N$ input ratios  by means of $N$ equations,
\begin{equation}
r(n)=\frac{b_{n+1}}{b_{n}}=r_N(n),\,\,0\le n \le N-1.
\end{equation}
It should be noted that this is a system of $N$ linear equations with $N$ unkowns ($q_m$ and $p_m$) which can be easily solved by  a computer. The hypergeometric ansatz is the crucial step that allows an extremely fast parameterization of large order Meijer-G approximants. 

\underline{\emph{Step 3: Hypergeometric approximants in the Borel plane.}} In this step we undertake the parameterization of hypergeometric approximants in the Borel plane. To do this for $N>1$, we use the calculated $p_m$ and $q_m$  to set two equations
\begin{eqnarray}
 \sum_{m=0}^l p_m x^m&=&0,\nonumber\\
1+\sum_{m=1}^l q_m y^m&=&0,
\end{eqnarray}
which yield two solution vectors $(x_1,\ldots,x_l)$ and $(y_1,\ldots,y_l)$. We refer to these vectors as \emph{hypergeometric vectors}. It follows from the definition of hypergeometric functions that the hypergeometric vectors determined in this way uniquely determine the hypergeometric function
\begin{equation}
B_{N}(\tau)\equiv \, _{l+1}F_{l}({\bf x},{\bf y},\frac{p_l}{q_l} \tau),
\end{equation}
where ${\bf x}=(1,-x_1,\ldots,-x_l)$, ${\bf y}=(-y_1,\ldots,-y_l)$ and $ _{l+1}F_{l}$ is a generalized hypergeometric function \cite{dlmf}. The function $B_N(\tau)$ is the hypergeometric approximant in the Borel plane; it provides an $N$-th order approximation to  the sum of the Borel-transformed series. Thanks to the hypergeometric ansatz and the hypergeometric vector equations, the function $B_N(\tau)$ can be easily parameterized for arbitrary large $N$.

\underline{\emph{Step 4: Meijer-G approximants.}} In this last step we need to reinstate the $n!$ removed from the expansion coefficients by means of the Borel transform. This is achieved by means of the  Laplace transform 
\begin{equation}
Z_{B,N}(g) \equiv \int_{0}^\infty e^{-\tau} B_{N}(g \tau) d\tau,
\end{equation}
which gives the desired approximantion  to the Borel sum of the asymptotic expansion of $Z(g)$ in the complexified $g$-plane. This expression admits the representation
\begin{equation}
Z_{B,N}(g)=\frac{\Pi_{i=1}^{l}\Gamma(-y_i)}{\Pi_{i=1}^{l}\Gamma(-x_i)}\MeijerG*{l+2}{1}{l+1}{l+2}{1,-y_1, \dots,-y_l}{1,1,-x_1, \dots, -x_l}{-\frac{q_l}{p_l g}},
\label{eq:MGF}
\end{equation}
where $\Gamma(x)$ is Euler's Gamma function and $\MeijerG*{m}{n}{p}{q}{a_1, \dots, a_p}{b_1, \dots, b_q}{z}$ is  Meijer's G-function (MGF)~\cite{MGF1,MGF2,dlmf}. This algorithm then transforms $N$ available input coefficients $z_n$ into a table of Meijer-G functions, which approximate the Borel sum of $Z(g)$. We once again emphasize how easily one can parameterize these extremely complex functions: {\it all what was needed was the hypergeometric ansatz and the resulting hypergeometric vectors}. Once these are determined the Meijer-G approximants can be parameterized for arbitrarily large orders.

There are various remarks that we would like to add before moving onto the practical application of this method . The even order approximants can be computed in exactly the same way, but one starts from a once-subtracted series, for instance from the once-subtracted  Borel-transformed series $(\sum_{n=0}^\infty b_n \tau^n-1)/(b_1 \tau)$ and the $B_N$ are calculated as above but with $l=N/2$. The  Laplace transform then gives the even order approximants.  It is also easy to compute approximations to the Mittag-Leffler (ML) sum \cite{Hardy1949}, which is a generalization of the Borel sum where one can remove superfactorial asymptotic coefficient growth (as opposed to factorial) by using a ML transform of the coefficients (as opposed to a Borel transform); in step 1 the ML-transformed coefficients are $b_n=a_n/\Gamma(\alpha n+\beta)$, where $\alpha$ and $\beta$ are real numbers, and the odd approximants in step 4 are given by $f_N(\lambda)=\int_0^\infty e^{-\tau} \tau^{\beta-1+\alpha} B_{N}(\lambda \tau^\alpha)d\tau$. When $\alpha=\beta=1$ ML summation is equivalent to Borel summation. These approximants also admit a MGF representation (not shown). Furthermore, further generalizations of  Borel and ML summation methods, where essentially arbitrary asymptotic coefficient growth is removed can be easily arrived at. 

It follows from the differential equation satisfied by MGFs that the  Meijer-G approximants given by  Eq.~\eqref{eq:MGF} provide a regularizing  analytic continuation of the divergent hypergeometric functions $_{l+2}F_l(-y_1, \dots,-y_l;1,1,-x_1, \dots, -x_l;\frac{p_l \lambda}{q_l})$~\cite{MGF1}. Such hypergeometric functions have zero radius of convergence; the fact that Meijer-G approximants are able to ``sum'' them clearly illustrates the potential of these functions for the summation of divergent series.

\section{Applications in Quantum Field Theory}\label{Sec:III}
In this section Meijer-G approximants are used to sum partition functions in QFT using only a few orders of perturbation theory. In particular, in Sec. III~A we will consider the summation of the partition function in $\phi^4$ theory, which is Borel summable. We will also consider the summation of non-Borel summable series by considering $\phi^4$ theory with degenerate minima~\cite{Marucho2008} in Sec. III~B, as well as a self-interacting QFT model in zero dimensions~\cite{Cherman2015} in Sec. III~C. In such cases the standard perturbation series needs to be upgraded to a transseries. These examples are often utilized for benchmarking new summation techniques, since they contain highly divergent series which are very difficult to sum. It will be shown that, remarkably, these cases constitute a best-case scenario for the application of the  Meijer-G summation technique developed above. This is so because the Meijer-G approximants converge---in all of these cases--- at order $N=5$. This means that by applying the Meijer-G summation procedure we arrived at a closed-form analytic expression; in other words, these partition functions belong to the space of functions reproducible by our summation technique and thus they are easily summable by means of Meijer-G approximants. The reason for this is that in these cases the hypergeometric ansatz turns out to be exact, and the Borel-transformed series sums exactly to a hypergeometric function of the form $_3F_2(1,h_1,h_2;h_3,h_4,h_5 \tau)$ [or, equivalently, to a hypergeometric function of the form $_2F_1(h_1,h_2,h_3,h_4\tau)$]. Therefore, in Sec.~III~D we complement our study by considering the summation of the transseries expansion for the double-well potential in quantum mechanics, including the one- and two-instanton contributions~\cite{JentschuraMultinstanton}, which constitutes a more challenging example since the approximants exhibit highly non-trivial convergence properties.
\subsection{Partition function in $\phi^4$ Theory}
The partition function in zero-dimensional $\phi^4$ theory is given by
\begin{equation}
Z(g)=\frac{1}{\sqrt{2\pi}}\int_0^\infty e^{\phi^2/2-g\phi^4/4!}d\phi,
\end{equation}
which for $\mathrm{Re}[g]>0$ can be written as
\begin{equation}
Z(g)=\sqrt{\frac{3}{2\pi g}}e^\frac{3}{4g}K_{\frac{1}{4}} \left( \frac{3}{4g} \right),
\end{equation}
where $K_\nu(x)$ is a modified Bessel function of the second kind. This partition function is commonly used to benchmark resummation techniques ---see Refs.~\onlinecite{Pollet2010, Garcia2017} for two recent examples. 

The first few terms of the asymptotic expansion about $g=0$ read
\begin{equation}
Z(g)\sim 1-\frac{1}{8}g+\frac{35}{384}g^2-\frac{385}{3072}g^3+\frac{25025}{98304}g^4+\cdots
\end{equation}
The expansion coefficients grow factorially at large orders and thus this expansion has zero radius of convergence.  Here the calculation of $Z(g)$ by direct resummation of the asymptotic expansion serves two main purposes. On the one hand it provides a simple test system for benchmarking against Borel-Pad\'e. On the other hand the parameterization of the approximants can be performed analytically at low orders by following our algorithm and thus is valuable from a didactical perspective.

Let us calculate the first-order and third-order Meijer G approximant for this problem analytically, by running the above-given algorithm explicitely. We start with the first order calculation and proceed step by step.
\begin{enumerate}
\item  \emph{Borel transform:} we calculate the Borel transformed coefficients $b_n=z_n/n!$.  In a first-order calculation we have just two coefficients, $b_0=1$ and $b_1=-1/8$. 
\item \emph{Hypergeometric ansatz:} we compute the ratios between consecutive coefficients as a function of $n$, $r(n)$ and we approximate the ratio by a rational function of $n$. In this case we have only one ratio $r(0)=b_1/b_0=-1/8$ and the rational function that approximates $r(n)$ is just a constant: $r(n)\approx r(0)$, $\forall n$. 
\item \emph{Hypergeometric approximant in the Borel plane:} in this case the hypergeometric vectors are empty, $(x_1,\ldots,x_l)=\{\}$ and $(y_1,\ldots,y_l)=\{\}$ and therefore $\mathbf{x}=1$ and $\mathbf{y}=\{\}$.  Since $l=(N-1)/2=0$ for $N=1$, the first order  hypergeometric approximant in the Borel plane is given by 
\begin{equation}
B_{H,N=1}(\tau)=\,_1F_0(1,r(0)\tau).
\end{equation}
This $_1F_0$ hypergeometric function is just the 0/1 Pad\'e approximant to the Borel-transformed series, i.e.,
\begin{equation}
B_{H,1}(\tau)=\frac{1}{1+r(0) \tau}.
\end{equation}
It should be clear that  the hypergeometric vectors are empty, and that the above-given hypergeometric function in the Borel plane reduces to the geometric case. Therefore our first-order hypergeometric approximant coincides with the 0/1 Pad\'e approximant. 

\item \emph{Meijer-G approximant:} once the hypergeometric approximant in the Borel plane is found, we can use Eq.~\ref{eq:MGF} to immediately write down the corresponding Meijer-G approximant in the complexified-$g$ plane, which reads
\begin{equation}
Z_{B,1}(g)=\frac{1}{8g}\MeijerG*{2}{1}{1}{2}{0}{0,0}{\frac{1}{8g}}.
\end{equation}
This Meijer-G approximant is just the 0/1 Borel-Pad\'e approximant. This is an interesting aspect of hypergeometric and Borel-hypergeometric resummation: to first order hypergeometric approximants are just 0/1 Pad\'e approximants; translating this observation to the Borel plane shows  that the corresponding Meijer-G  (Borel-hypergeometric) approximant is just the 0/1 Borel-Pad\'e approximant. Indeed,
\begin{equation}
Z_{B,1}(g)=\int_0^\infty \frac{e^{-\tau}}{1+r(0)\tau} d\tau=-\frac{r(0)}{g}U\left(1,1,\frac{-r(0)}{g}\right),
\end{equation}
is both the 0/1 Borel-Pad\'e approximant and the first order Meijer-G approximant.
\end{enumerate}
The conclusion from this first example calculation is that our first-order hypergeometric-Borel (Meijer-G) approximant is  the first-order Borel-Pad\'e approximant. This illustrates a general property of Meijer-G approximations to the Borel sum: their first order is just identical to a first-order Borel-Pad\'e approximant. The interested reader can now look to the closely related approach put forward in Ref.~\onlinecite{Garcia2017} ---both approaches take very different routes beyond first order.

Next we run again our algorithm to obtain the third order approximant. While the first order Meijer-G approximant  is identical to the first order Borel-Pad\'e approximant we will see shortly that the third order Meijer-G approximant is substantially more accurate than Borel-Pade approximants of the same and much higher orders.  
\begin{enumerate}
\item  \emph{Borel transform}. In this case  we have four Borel-transformed coefficients: $b_0=1$, $b_1=-1/8$, $b_2=35/768$ and $b_3=-385/18432$.
\item \emph{Hypergeometric Ansatz}. Here we have three ratios: $r(0)=-1/8$, $r(1)=-35/96$, and $r(3)=-11/24$. We approximate $r(n)$ as $r(n)=r_3(n)$ where 
\begin{equation}
r_3(n)= \frac{p_0+p_1 n}{1+q_1 n},
\end{equation}
and use the known ratios, $r(0)$, $r(1)$ and $r(2)$ to find $p_0$, $p_1$ and $q_1$ by requiring 
\begin{equation}
r(n)=r_3(n),\qquad n=0,1,2,
\end{equation}
which leads to three equations
\begin{eqnarray}
r(0)&=&-\frac{1}{8}=p_0,\\
r(1)&=&-\frac{35}{96}=\frac{p_0+p_1}{1+q_1},\\
r(2)&=&-\frac{11}{24}=\frac{p_0+2p_1}{1+2q_1},
\end{eqnarray}
with three unknowns, $p_0$, $p_1$ and $q_1$.
These equations yield a solution
\begin{equation}
p_0=-\frac{1}{8},\,p_1=-\frac{113}{216},\,q_1=\frac{7}{9}.
\end{equation}
Therefore our third order rational approximation to the ratios $r(n)$ reads
\begin{equation}
r(n)\approx r_3(n)=\frac{-1/8-113 n/216}{1+7n/9}.
\end{equation}
\item \emph{Hypergeometric Approximant in the Borel plane}. To build the hypergeometric approximants in the Borel plane we need to find the hypergeometric vectors, that is to  find the values of $x$ and $y$ that solve these two equations
\begin{eqnarray}
p_0+p_1 x&=&0,\\
1+q_1 y &=&0,
\end{eqnarray}
which yield $x_1=-p_0/p_1$ and $y_1=-1/q_1$ and thus find the vectors $\mathbf{x}=(1,p_0/p_1)$ and $\mathbf{y}=(-1/q_1)$. Hence the third order hypergeometric approximant in the Borel plane is
\begin{equation}
B_{H,3}(\tau)=\,_2F_1(1,\frac{p_0}{p_1},\frac{1}{q_1},\frac{p_1}{q_1}\tau).
\end{equation}
Substituting values of $p_i$ and $q_i$ we get
\begin{equation}
B_{H,3}(\tau)=\,_2F_1(1,\frac{27}{113},\frac{9}{7},-\frac{113}{168}\tau).
\end{equation}
\item \emph{Meijer-G approximant.} We read-off the Meijer-G approximant directly from the hypergeometric vectors to get a third order approximant that can be compactly written as
\begin{equation}
Z_{B,3}(g)=\frac{\Gamma(9/7)}{\Gamma(27/113)}\MeijerG*{3}{1}{2}{3}{1,9/7}{1,1,27/113}{\frac{168}{113 g}}.
\end{equation}
\end{enumerate}
\begin{figure}
\center
\includegraphics[width=0.5\textwidth]{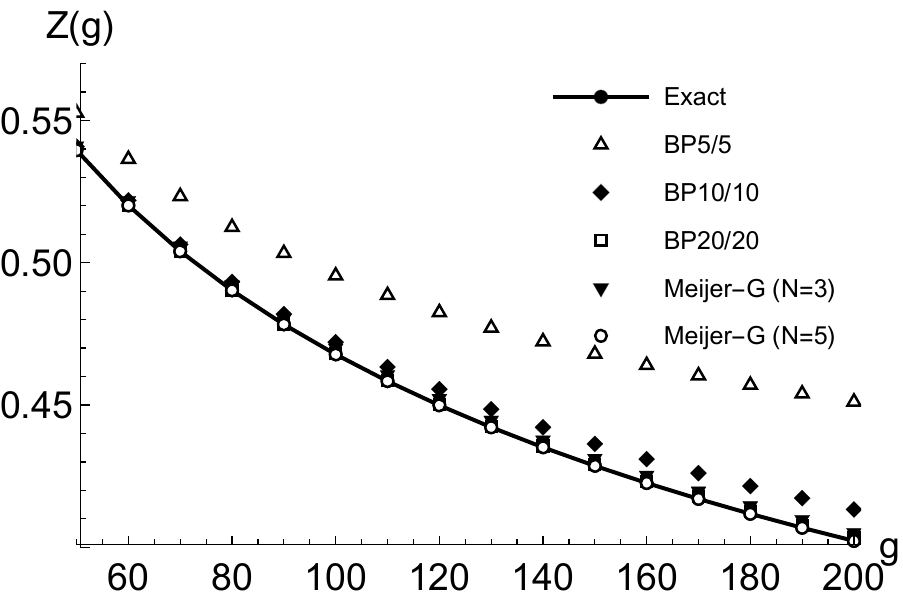}
\caption{$Z(g)$ for 0-dimensional $\phi^4$ theory calculated using MGFs and Borel-Pad\'e approximants. For large values of $g$ the third-order Meijer-G approximant (filled inverted triangles) and the fifth order Mejer-G approximant (empty circles) are  compared with higher-order 5/5 (empty triangles), 10/10 (filled diamonds) and 20/20 (empty squares)   Borel-Pad\'e approximants  and with the exact $Z(g)$ (solid line with filled circles). The third order Meijer-G approximant is more accurate that 5/5 and 10/10 Borel-Pad\'e approximants, and slightly less accurate than the 20/20 Borel-Pad\'e approximant. The fifth-order Meijer-G approximant is exact and thus more accurate than any Borel-Pad\'e approximant.}
\label{Fig:fig2}
\end{figure}
\begin{table*}[t]
\begin{center}
 \begin{tabular}{||c |c| c|c||} 
 \hline
 Method  & $g=-1$  & $g=-10$ & $g=-100$ \\ [0.5ex] 
 \hline\hline
BP2/2&$1.132752 - 0.129446\mathrm{i}$&$0.598308 - 0.424956 \mathrm{i}$&$0.473216 - 0.070069\mathrm{i}$\\
\hline
 BP5/5 &$1.133180 - 0.144446\mathrm{i}$  &$0.784563 - 0.458166\mathrm{i}$&$ 0.300204 - 0.251866 \mathrm{i}$\\ 
 \hline
 BP10/10 & $1.133022-0.144983\mathrm{i}$  &$ 0.740363 - 0.458776 \mathrm{i}$&$0.329910 - 0.450161  \mathrm{i}$\\
 \hline
 BP20/20 &$1.133028- 0.144995\mathrm{i}$ & $0.746175 - 0.494474\mathrm{i}$&$0.402820 - 0.519661 \mathrm{i}$ \\
 \hline\hline
 Meijer-G ($N=3$) & $1.133285- 0.144952\mathrm{i} $ &$0.744345 - 0.436217\mathrm{i}$&$0.386356 - 0.321210 \mathrm{i}$ \\
 \hline
 Meijer-G ($N=5$) & $1.133029 - 0.144984\mathrm{i}$ &$0.746390 - 0.436845\mathrm{i}$ &$0.384675 - 0.325851 \mathrm{i}$\\
 \hline\hline
 Exact &$1.133029 - 0.144984\mathrm{i}$ &$0.746390 - 0.436845\mathrm{i}$ &$0.384675 - 0.325851 \mathrm{i}$\\ 
 \hline\hline
\end{tabular}
\end{center}
\caption{$Z(g)$ for 0-dimensional $\phi^4$ theory evaluated ``on the cut'', for  $g=-1$, $g=-10$ and $g=-100$ using Borel-Pad\'e approximants of orders $4$ (BP2/2), $10$ (BP5/5), $20$ (BP10/10) and $40$ (BP20/20), compared with  third and fifth order Meijer-G approximants and with the exact value. All resuts are given to six significant digits. The accuracy of the third order Meijer-G approximant is comparable to that of BP approximants of higher order at weak couplings ($g=-1$) and greater at large couplings ($g=-10$ and $g=-100$). The fifth order Meijer-G approximant is exact.}
\label{Tab:table1}
\end{table*}
One can easily calculate higher order Meijer-G approximants. It turns out that the fifth order Meijer-G approximant is converged and equal to the exact Borel sum, i.e.,
\begin{equation}
Z_{B,5}(g)=Z(g),
\end{equation}
and all higher order Meijer-G approximants are also equal to $Z(g)$, i.e.,
\begin{equation}
Z_{B,5}(g)=Z_{B,7}(g)=\cdots=Z(g).
\end{equation}
What is happening is that the rational approximations used in the hypergeometric ansatz have converged at fifth order, i.e., the ratio between consecutive Borel-transformed coefficients is indeed a rational function of $n$, and, in fact, of the form
\begin{equation}
r(n)=\frac{p_0+p_1 n + p_2 n^2}{1+q_1 n + q_2 n^2},
\end{equation}
specifically
\begin{equation}
r(n)=\frac{-1/8-2 n/3 -2  n^2/3}{1+2 n +  n^2},
\end{equation}
which reproduces the ratios between Borel-transformed coefficients up to arbitrarily high orders. Approximating these ratios by  rational functions of higher order, such as $r_7(n)$, one finds the same rational function once again. Hence the approximants of order fiver or higher are converged.

We now compare the performance of Meijer-G approximants with that of Borel-Pad\'e approximants. In Fig.~\ref{Fig:fig2} we compare the $5/5$, $10/10$ and $20/20$ Borel-Pad\'e approximants (of orders $10$, $20$ and $40$ respectively) with the third and fifth order Meijer-G approximant for large values of $g$. It is clear that the third order Meijer-G approximant is more accurate than the 10/10 Borel-Pad\'e approximant, but less accurate than the $40$-order 20/20 Borel-Pad\'e approximant.  The fifth order Meijer-G approximant is exact and it is therefore more accurate than any Borel-Pad\'e approximant. 

It is instructive to compare $Z(g)$ with $Z_3(g)$ for $g<0$. For instance
\begin{equation}
Z(-10+\mathrm{i}\epsilon)=0.7463895836 - \mathrm{i}0.4368446698,
\end{equation}
while
\begin{equation}
Z_3(-10+\mathrm{i}\epsilon)=0.7443450750 - \mathrm{i}0.4362172724,
\end{equation}
which demonstrates the great accuracy of Meijer-G approximants for rather large negative couplings; in particular the imaginary part of $Z(g)$ is reproduced within a percent.  In Table~\ref{Tab:table1} we compare again the third and fifth order order Meijer-G approximants  against the Borel-Pad\'e approximants---this time for negative couplings. We see that for $g=-1$  all approximants are very accurate. The third order Meijer-G approximant is more accurate than the 2/2 Borel-Pad\'e approximant, but less accurate than the other Borel-Pad\'e approximants shown. For $g=-10$ the Meijer-G approximant is already more accurate than 2/2, 5/5 and 10/10 Borel-Pad\'e approximants. Finally for $g=-100$ the Meijer-G approximant is more accurate than all the Borel-Pad\'e approximants reported.  The fifth order Meijer-G approximant reproduces the exact result.


These findings demonstrate how to easily build accurate Meijer-G approximations to the sum of a Borel-summable divergent series, and are confirmed by calculations for the quartic anharmonic oscillator and the Heisenberg-Euler action for QED, which will be shown elsewhere. These approximants have great potential for the summation of Borel-summable series and should be accurate in cases where the convergence of the Borel-transformed series is limited by a branch point singularity. Here we have considered the case of 0-dimensional $\phi^4$ theory, which is a challenging example for standard approximants ---which typically require a large number of coefficients in order to yield accurate estimates for $Z(g)$; in contrast Meijer-G approximants converge rapidly to the exact answer.  In many cases, however, perturbation expansions are not Borel-summable. In such cases one needs to sum a resurgent transseries. Below it is shown that Meijer-G approximants can also be used to provide economical and accurate approximations to the ``sum'' of  such expansions.

\subsection{Degenerate Vacua}
Marucho~\cite{Marucho2008} considered a partition function of the form
\begin{equation}
Z(g)=\frac{1}{\sqrt{2\pi}}\int_{-\infty}^{\infty}e^{-\phi^2(1-\sqrt{g}\phi)^2/2}d\phi, \qquad g>0.
\label{eq:maruchoz}
\end{equation}
The term $\phi^2/2$ in the exponent may be regarded as a free-field ($g=0$) and a traditional perturbative approach, such as diagrammatics, results in an asymptotic expansion in powers of $g$. The first few terms of this expansion are
\begin{equation}
Z(g)\sim1+6g+210g^2+13860g^3+\cdots,
\label{eq:asym}
\end{equation}
and its general term  can be found in the paper  in Ref.~\onlinecite{Marucho2008}. Hence we are once again in a situation where we happen to know all of the coefficients. We stress that this only happens in toy models and expansion coefficients are rarely known at large orders.  With these coefficients we can easily run our algorithm and find the odd order Meijer-G approximants, which read
\begin{eqnarray}
Z_{B,3}(g)&=& -\frac{0.00733}{g}\MeijerG*{3}{1}{2}{3}{0,0.286}{-0.761,0,0}{-\frac{0.0310}{g}},\nonumber\\
Z_{B,5}(g)&=&-\frac{0.00701}{g}\MeijerG*{4}{1}{3}{4}{0,0,0}{-0.25,-0.75,0,0}{-\frac{0.0312}{g}},\nonumber\\
Z_{B,7}(g)&=&-\frac{0.00701}{g}\MeijerG*{5}{1}{4}{3}{0,0,0,0}{-0.25,-0.75,0,0,0}{-\frac{0.0312}{g}}.\nonumber \\
&\vdots&\nonumber
\end{eqnarray}
These Meijer-G approximants to the Borel sum are fully converged  at order  $N=5$ and larger; by increasing the order of the approximants we get the same function again and again:  $Z_{B,5}=Z_{B,7}=\cdots$. This means that we have found the exact Borel sum by means of our approach; \emph{the Borel sum of} $Z(g)$ belongs to the space of reproducible functions associated with Borel-Hypergeometric resummation. The converged approximants agree with the exact Borel sum~\cite{Marucho2008} given by
\begin{equation}
Z_B(g)\approx\frac{\mathrm{i}0.1e^{-0.0156/g}}{\sqrt{x}}K_{-1/4}\left(-\frac{0.0156}{g}\right).
\label{eq:bsum}
\end{equation}
Our approximants actually find the exact Borel sum for this problem with just a few orders. In contrast Borel-Pad\'e would have required very high order coefficients to produce accurate results up to modest values of the coupling.  Unfortunately---for the present problem---having very accurate approximations to the Borel sum turns out to be  insufficient  to accurately approximate $Z(g)$.
\begin{figure*}
\center
\includegraphics[width=0.45\textwidth]{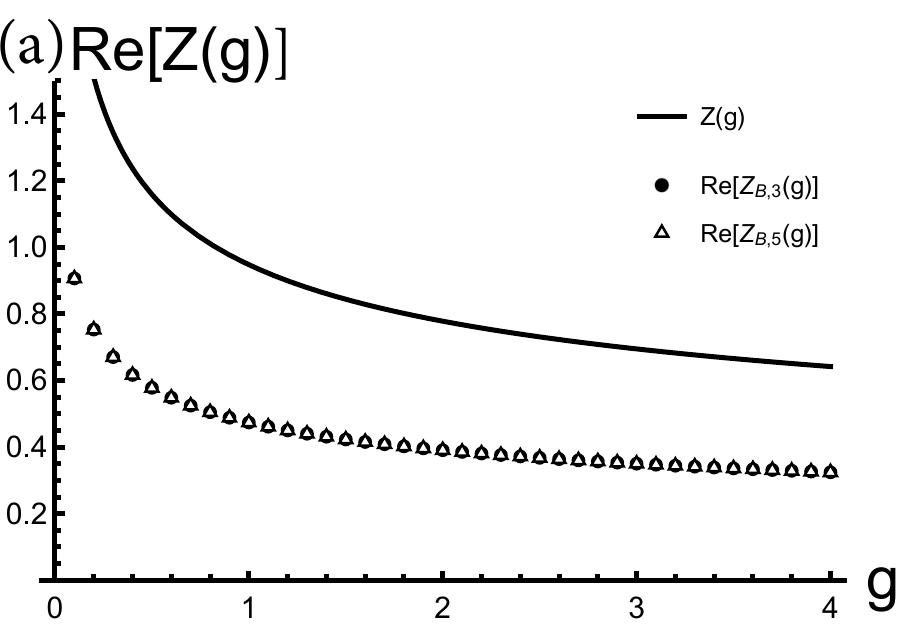}
\includegraphics[width=0.45\textwidth]{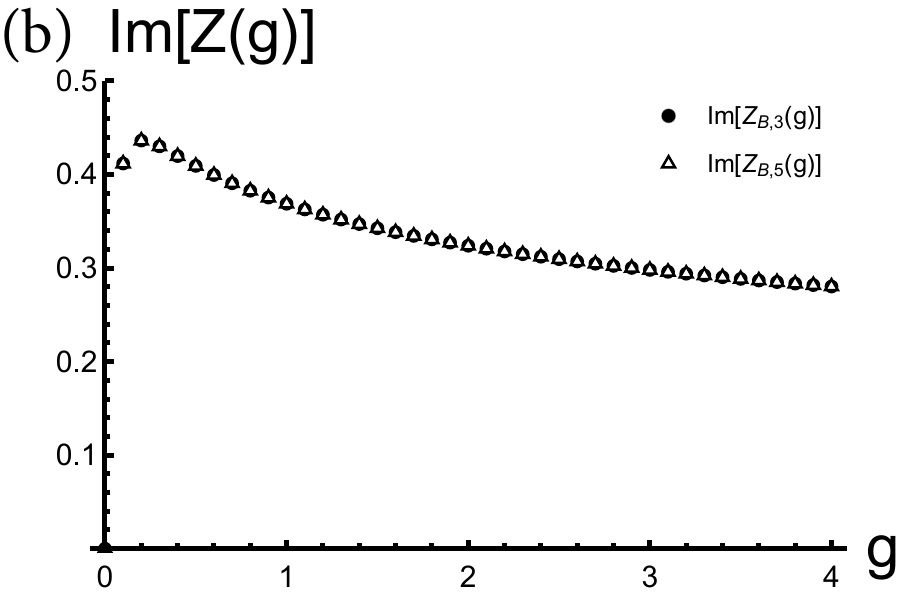}
\caption{Partition function, $Z(g)$, for $\phi^4$ theory with degenerate minima.  (a) The real part of the exact $Z(g)$ (solid) is compared with the real part of the third- ($Z_{B,3}(g)$, filled circles) and fifth-order ($Z_{B,5}(g)$, empty triangles) Meijer-G approximants to the Borel sum of $Z(g)$. The Meijer-G approximants are well converged for the range of values of $g$ shown and are not clearly distinguishable on the scale of the plot. The fifth order approximant is converged to the exact Borel sum given by Marucho \cite{Marucho2008}. The real part of the Borel sum underestimates $Z(g)$ by a factor of two , $\mathrm{Re}[Z_{B,5}(g)]=Z(g)/2$. (b) Imaginary parts of the Meijer-G approximants are not zero, while $Z(g)$ is manifestly real,  $\mathrm{Im}[Z_{B,5}(g)]\neq 0$ while  $\mathrm{Im}[Z(g)]= 0$.  Perturbation theory for $\phi^4$ with degenerate minima is not Borel-summable. }
\label{Fig:fig3}
\end{figure*}
The problem is of course that $Z(g)$ is not Borel-summable.  This is clearly seen in Figure~\ref{Fig:fig3} which shows the real and imaginary parts of $Z_{B,3}(g)$  and $Z_{B,5}(g)$   together with the exact $Z(g)$.  The Borel sum fails to accurately represent $Z(g)$; in Fig.~\ref{Fig:fig3}(a) we see that the real part of the Borel sum is off by a constant factor of 2;  in Fig.~\ref{Fig:fig3}(b) we see that the Borel sum posseses an imaginary part, while  $Z(g)$ is manifestly real. Note that the $N=3$ approximant is nearly converged while the $N=5$ approximant is fully converged to the exact value of the Borel sum, given in Ref.~\onlinecite{Marucho2008}. 

\begin{figure}[t!]
\center
\includegraphics[width=0.45\textwidth]{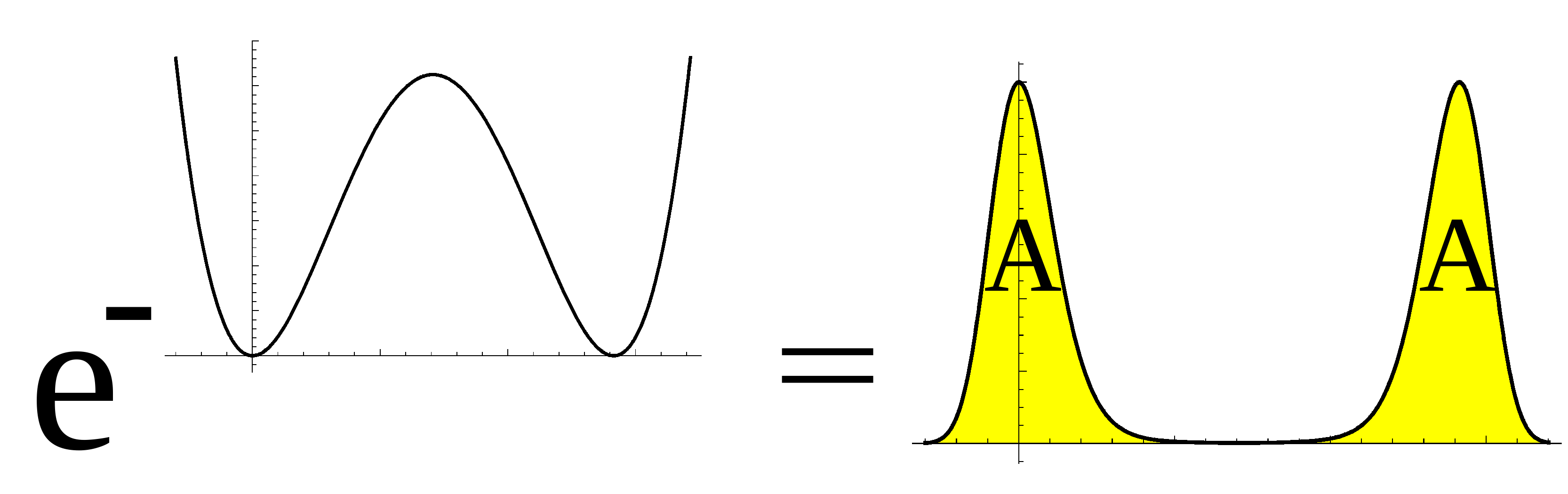}
\caption{Summability problem in $\phi^4$ theory with degenerate minima. Exponentiating a symmetric double-well with $g=0.005$ yields two equal-area gaussians, each centered on its respective well. The left well is called perturbative saddle and the right well is called non-perturbative saddle. For any $g>0$, no matter how small, $Z(g)$ is the total area under the gaussians. Borel summation of perturbation theory yields exactly $Z(g)/2$ plus an imaginary part called the non-perturbative ambiguity, which is exponentially supressed as $g\rightarrow 0$. The Borel sum of the standard divergent perturbation theory of $Z(g)$ accounts only for contributions coming from the perturbative saddle.}
\label{Fig:fig4}
\end{figure}

When performing a Borel summation it is not uncommon to obtain imaginary parts.  While in some cases it is rather straightforward to give them a physical meaning, in the present case it is not so clear and furthermore such physical interpretation is not our objective---which is to accurately  approximate $Z(g)$.  This imaginary part is called non-perturbative ambiguity. It is ambiguous because it changes sign depending on whether one evaluates $Z_B(g+\mathrm{i}\epsilon)$ or $Z_B(g-\mathrm{i}\epsilon)$, i.e.  $\mathrm{Im}[Z_B(g+\mathrm{i}\epsilon)]=-\mathrm{Im}[Z_B(g-\mathrm{i}\epsilon)]$. Non-Borel summability for this problem  is  explained in Ref.~\onlinecite{Marucho2008}.  For completeness  we give our version  here (see Fig.~\ref{Fig:fig4}).  For very small but non-zero couplings the integrand in $Z(g)$ contains contributions from two identical gaussians, each of which sits on a different saddle, one at $x=0$ 
and the other at $x=1/\sqrt{g}$. In the literature the former is referred to as the ``perturbative saddle'' while the latter is known as the ``non-perturbative saddle''. If the area under each gaussian is  $A$, then $Z(g\rightarrow0^+)=2A$. When doing perturbation theory around $g=0$ one is effectively taking into account contributions only from the perturbative saddle and thus the perturbative estimate will be $Z(g\rightarrow0^+)=A$, which is wrong by a factor of two. The Meijer-G approximants  succeed in summing up the perturbation expansion around $g=0$ from only a few terms, but that is not enough. The Borel sum misses the contribution from the 
second gaussian entirely and thus underestimates the real part of $Z(g)$ by a factor of two. It is  then clear that $Z(g)$ has an essential singularity at $g=0$. Perturbation theory is not sufficient to accurately evaluate $Z(g)$ even in the limit $g\rightarrow0$.
\begin{figure}
\center
\includegraphics[width=0.45\textwidth]{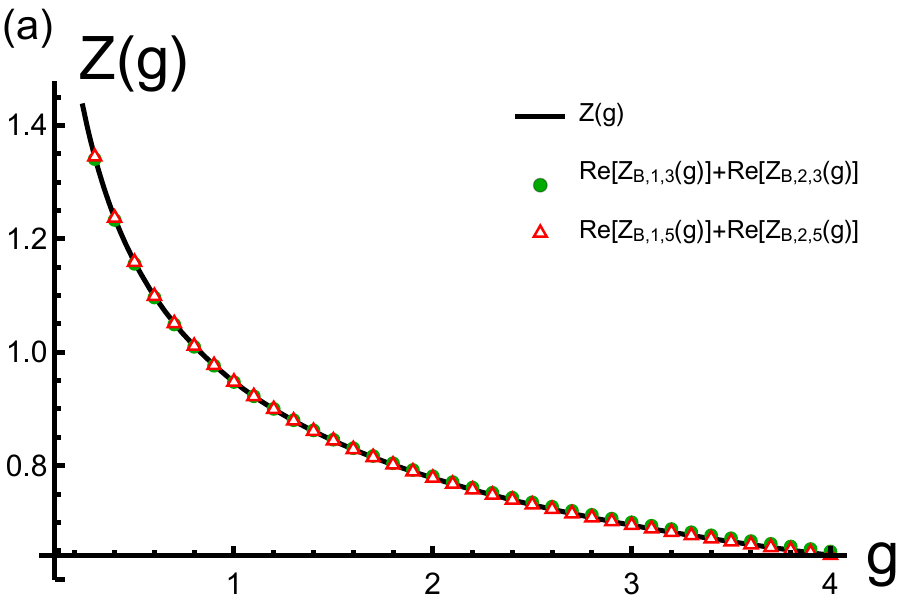}
\includegraphics[width=0.45\textwidth]{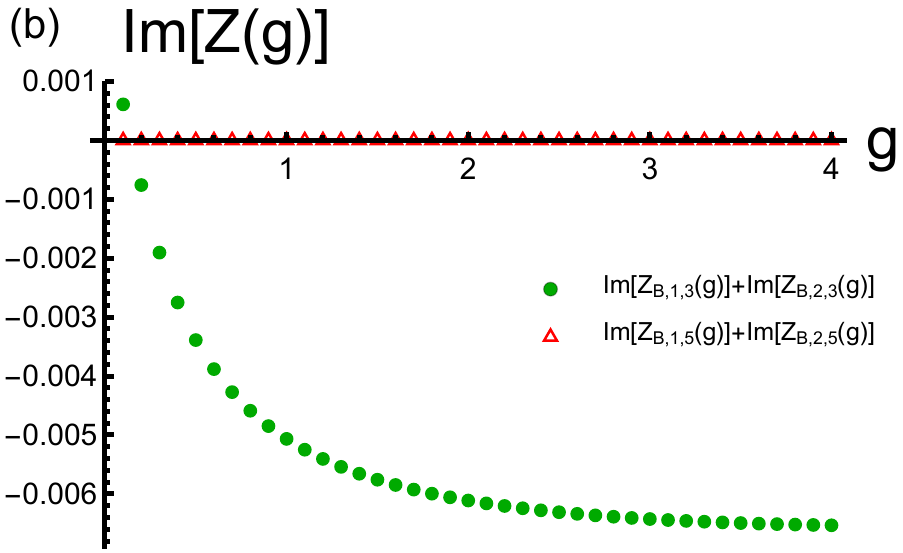}
\caption{Transeries summation by Meijer-G approximants for $\phi^4$ theory with degenerate minima. (a)   Exact real part of $Z(g)$ versus third- and fifth-order summed-up transseries. 
(b) Imaginary part of $Z(g)$ as computed by third order and fifth order Meijer-G approximants; the third order  Meijer-G approximants nearly cancel the non-perturbtaive ambiguity (compare with Fig.~\ref{Fig:fig3}(b)); the fifth order approximants are exact and thus have zero imaginary part.  }
\label{Fig:fig5}
\end{figure}

In order to obtain an accurate estimate of $Z(g)$ we need to upgrade the perturbation expansion to a resurgent transseries which, for the present problem, is of the form
\begin{widetext}
\begin{eqnarray}
Z(g)&=&\pm \mathrm{i}\sqrt{g}e^{-\frac{1}{32g}}\,_2F_0(1/4,3/4,-32g)+2\,_2F_0(1/4,3/4,32g)\label{eq:transmarucho}\\
&=&\pm \mathrm{i}\sqrt{g}e^{-\frac{1}{32g}}(1-6g+210g^2-\cdots)+2(1+6g+210g^2+\cdots), \nonumber
\end{eqnarray}
\end{widetext}
where the $_2F_0$ factors are divergent hypergeometric series and the upper sign is for $\mathrm{Im}\, g>0$, while the lower sign is for  $\mathrm{Im}\, g<0$. This transseries contains two divergent series that need to be summed. The so-called Borel-Pad\'e-\'Ecalle method uses the  Borel-Pad\'e technique to sum each of these divergent series \cite{Dingle1973, Berry1991, Aniceto2015a, Aniceto2015b, Cherman2015, Marino2015}. Here we use instead Meijer-G approximants to sum each of the divergent series that appear in the transseries expansion. Such an approach can then be described as the Borel-Hypergeometric-\'Ecalle summation method. Meijer-G approximants are extremely-well suited to sum this expansion exactly since---as discussed in in Section~\ref{Sec:II}---they are regularizing analytic continuations of the divergent hypergeometric series  $_{n+2}F_n$.   Therefore the \emph{exact}  Borel sum of each of the two divergent hypergeometric functions that enter Eq.\eqref{eq:transmarucho}   can be obtained directly from Meijer-G approximants of order $N\ge5$.   Defining
\begin{eqnarray}
Z_1(g)&\equiv& \pm \mathrm{i}\sqrt{g}e^{-\frac{1}{32g}}\,_2F_0(1/4,3/4,-32g),\\
Z_2(g)&\equiv& 2\,_2F_0(1/4,3/4,32g),
\end{eqnarray}
we can find their respective $N$th order Meijer-G approximations to their Borel sums, denoted $Z_{B,1,N}$ and $Z_{B,2,N}$. The third order approximants are 
\begin{eqnarray}
Z_{B,1,3}(g)&=&\pm e^{-\frac{1}{32g}} \frac{\mathrm{i}0.00733}{\sqrt{g}}\MeijerG*{3}{1}{2}{3}{0,0.286}{-0.761,0,0}{\frac{0.0310}{g}},\\
Z_{B,2,3}(g)&=&-2 \frac{0.00733}{g}\MeijerG*{3}{1}{2}{3}{0,0.286}{-0.761,0,0}{-\frac{0.0310}{g}},
\end{eqnarray}
and the third order approximation to the full transseries is
\begin{table*}[t]
\begin{center}
 \begin{tabular}{||c |c| c|c||} 
 \hline
Approximant  & $g=1$  & $g=10$ & $g=100$ \\ [0.5ex] 
 \hline\hline
$Z_{B,3}(g)$&$0.4737936480 + 0.3687242092  \mathrm{i}$&$0.2556937980 + 0.2286095678 \mathrm{i}$&$0.1444897734 + 0.1335395026 \mathrm{i}$\\
\hline
$Z_{B,5}(g)$&$0.4739795910 + 0.3727956186 \mathrm{i}$&$0.2505643011 + 0.2323170269  \mathrm{i}$&$0.1375877371 + 0.1343382972\mathrm{i}$\\
\hline\hline
$Z_{B,1,3}(g)+Z_{B,2,3}(g)$&$0.9475872960 - 0.0101327393 \mathrm{i}$&$0.5113875961 - 0.0127576878 \mathrm{i}$&$0.2889795468 - 0.0090959809 \mathrm{i}$\\
\hline
$Z_{B,1,5}(g)+Z_{B,2,5}(g)$&$0.9479591819$  &$0.5011286021$&$0.2751754743$\\ 
 \hline\hline
 Exact &$0.9479591819$ &$0.5011286021$ &$0.2751754743$\\ 
 \hline\hline
\end{tabular}
\end{center}
\caption{$Z(g)$ for 0-dimensional $\phi^4$ theory with degenerate minima evaluated for  $g=1$, $g=10$ and $g=100$ using the third and fifth order Meijer-G approximants to the Borel sum, $Z_{B,3}(g)$ and $Z_{B,5}(g)$, as well as the third and fifth order Meijer-G approximations to the resurgent transseries, $Z_{B,1,3}(g)+Z_{B,2,3}(g)$ and $Z_{B,1,5}(g)+Z_{B,2,5}(g)$.  We also include the exact result obtained by evaluating $Z(g)$ numerically in Eq.~\eqref{eq:maruchoz}. The approximations to the Borel sum fail to reproduce both the real part and the imaginary part of the exact result. In contrast the Meijer-G resummation of the resurgent  transseries for this problem gives a better account of  the exact $Z(g)$: the third-order Meijer-G approximant is in excellent agreement with the exact result, nearly cancelling the non-perturbative ambiguity; the fifth order Meijer-G approximant is exact. All quantities are given to 12 significant digits.}
\label{Tab:table2}
\end{table*}
\begin{equation}
Z(g)=Z_{B,1,3}(g)+Z_{B,2,3}(g)+O(g^4).
\end{equation}
while the the fifth order approximant is converged to the exact value. The fifth order approximantion to the transseries actually reconstructs $Z(g)$ {\it exactly}:
\begin{equation}
Z(g)=Z_{B,1,5}(g)+Z_{B,2,5}(g).
\end{equation}

Hence we use Meijer-G functions to sum up the $_2F_0$ divergent hypergeometric series that appear in the transseries given by Eq.\eqref{eq:transmarucho}. In Figure \ref{Fig:fig5} we compare the exact $Z(g)$ with the results of third- and fifth-order Borel-Hypergeometric-\'Ecalle summation. In Fig.~\ref{Fig:fig5}(a) we see that third order and fifth order Meijer-G approximants are excellent approximations to the exact $\mathrm{Re} Z(g)$; the factor of two discrepancy has been removed by proper transseries summation. In Fig.~\ref{Fig:fig5}(b) we see that the non-perturbative ambiguity has also been cancelled; while for third-order Meijer-G approximants this cancellation is not complete the size of the ambiguity has been reduced by two orders of magnitude, as can be seen by comparing Fig.~\ref{Fig:fig5}(b) with Fig.~\ref{Fig:fig3}(b). The  cancellation is exact for fifth-order Meijer-G approximants that sum the transseries exactly. Table II shows both the non-perturbative ambiguity and the missing factor 2 in the Meijer-G approximants to the Borel sum of $Z(g)$ ($Z_{B,3}$ and $Z_{B,5}$) for selected values of $g$ (ranging from intermediate to very strong couplings). When the Borel-Hypergeometric-\'Ecalle summation is adopted the ambiguity is removed and the factor 2 is restored. The third order approximant, while not exact, is remarkably accurate.

Using the integral considered by Marucho \cite{Marucho2008} we have illustrated the application of Meijer-G approximants to transseries summation. Such approach can be thought of as Borel-Hypergeometric-\'Ecalle summation, as opposed to Borel-Pad\'e-\'Ecalle and consists in using Meijer-G approximants (or hypergeometric approximants on the Borel plane; or rational approximations for the ratio between consecutive coefficients) to sum the divergent series that contribute to the resurgent transseries. In the present case we found that the transseries can be summed exactly with fifth order Meijer-G approximants. While not exact, the third order  approximation provides excellent approximations to $Z(g)$ and removes most of the non-perturbative ambiguity. 
\subsection{Self-interacting QFT}
Ref.~\onlinecite{Cherman2015} considers the following partition function of a ``0+0-dimensional self-interacting QFT''
\begin{equation}
Z(g)=\frac{1}{\sqrt{2\pi g} }\int_{-\pi/2}^{\pi/2} e^{\frac{\sin(\phi)^2}{2g}}d\phi=\frac{\pi}{\sqrt{g}}e^{-\frac{1}{4g}}I_0\left(\frac{1}{4g}\right),
\end{equation}
where $I_0(x)$ is a modified Bessel function of the first kind. $Z(g)$ has the asymptotic expansion
\begin{equation}
Z(g)\sim1+\frac{g}{2}+\frac{9g^2}{8}+\frac{75 g^3}{16}+\frac{3675g^4}{128}+\cdots
\end{equation}
Using the techniques put forward in Section II we once again obtain a convergent set of Meijer-G approximantions to the Borel sum of this series. The converged Meijer-G approximant to the Borel sum is  of fifth order and reads
\begin{eqnarray}
Z_{B,5}(g)&=&- \frac{1}{4\pi g}\MeijerG*{4}{1}{3}{4}{0,0,0}{-1/2,-1/2,0,0}{-\frac{1}{4g}}\\
&=& -\frac{\mathrm{i}}{2\pi g}e^{-\frac{1}{4g}}K_0\left(-\frac{1}{4g}\right).\nonumber
\end{eqnarray}
As in the previous example, increasing the order of the approximant yields the same Meijer-G function. The converged approximants can be compared with the very accurate, but not exact, third-order Meijer-G approximant
\begin{equation}
Z_{B,3}(g)=-\frac{26}{53 g}\frac{\Gamma(\frac{19}{13})}{\Gamma(\frac{19}{53})}\MeijerG*{3}{1}{2}{3}{0,6/13}{-34/53,0,0}{-\frac{26}{53 g}}.\\
\end{equation}
Evaluating $Z(g)$ and $Z_{B,5}(g)$ numerically for selected values of $g$ we see that 
\begin{equation}
\mathrm{Re}[Z_{B,5}(g)]=\mathrm{Re}[Z(g)], 
\end{equation}
but
\begin{equation}
\mathrm{Im}[Z_{B,5}(g)]\neq 0, 
\end{equation}
while
\begin{equation}
\mathrm{Im}[Z(g)]= 0. 
\end{equation}
Thus we find the non-perturbative ambiguity once again. For example, evaluating the third- and fifth-order approximants at $g=1$ we obtain
\begin{eqnarray}
Z_{B,3}(g=1\pm\mathrm{i}\epsilon)&=&0.9903122408877890894\ldots  \nonumber\\
&\pm&   \mathrm{i} 0.4813082375368570801\ldots , \nonumber\\ \nonumber
Z_{B,5}(g=1\pm\mathrm{i}\epsilon)&=&0.9913929921688975613\ldots \\ \nonumber
&\pm&  \mathrm{i} 0.4789408454106600542\ldots, \nonumber
\end{eqnarray}
while 
\begin{equation}
Z(g=1)=0.9913929921688975613\ldots
\end{equation}
Repeating at the very large value of $g=100$ we find
\begin{eqnarray}
Z_{B,3}(g=100\pm\mathrm{i}\epsilon)&=&0.13677671640883210679\ldots \nonumber \\
&\pm&   \mathrm{i} 0.23483780883795517888 \ldots  , \nonumber\\
Z_{B,5}(g=100\pm\mathrm{i}\epsilon)&=&0.12501867187315494524\ldots  \nonumber\\
&\pm&  \mathrm{i} 0.24304192933460168829 \ldots ,\nonumber 
\end{eqnarray}
while 
\begin{equation}
Z(g=100)=0.12501867187315494524\ldots
\end{equation}

\begin{figure*}
\center
\includegraphics[width=\textwidth]{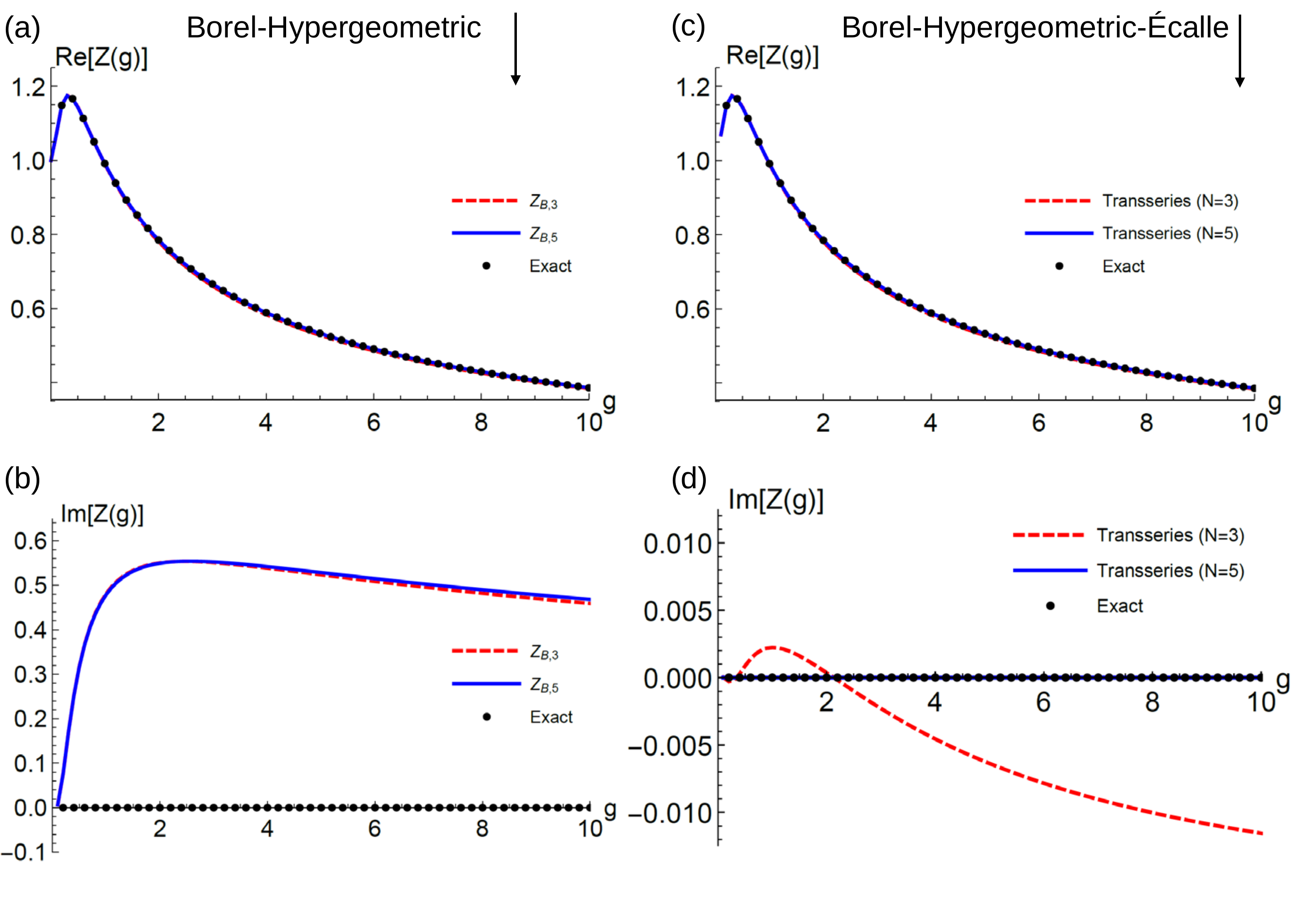}
\caption{$Z(g)$ for self-interacting 0-dimensional QFT as a function of $g$  calculated using Meijer-G approximants to the Borel sum (Borel-Hypergeometric summation; left panels) and to the Borel-\'Ecalle sum (Borel-Hypergeometric-\'Ecalle sum, right panels), and compared with the exact result (dots). (a) Real part of the Borel sum calculated using Meijer-G approximants of order three (dashed) and five (solid), compared to the exact result; the approximants return excellent approximations to $\mathrm{Re}[Z(g)]$. (b) As in (a), but for $\mathrm{Im}[Z(g)]$; while the Meijer-G approximants are very nearly converged for the range of values of $g$ shown, they fail to reproduce the exact value, $\mathrm{Im}[Z(g)]=0$. (c) As in (a), but this time we use Meijer-G approximants of third order (dashed) and fifth order (solid) to approximate the Borel-\'Ecalle sum, obtaining once again excellent agreement with the exact $\mathrm{Re}[Z(g)]$. (d)  As in (c)  but for  $\mathrm{Im}[Z(g)]$; the combination of Meijer-G approximants with the Borel-\'Ecalle approach results in a nearly complete cancellation of the non-perturbative ambiguity at third order---the cancellation is complete for $N=5$.}
\label{Fig:fig6}
\end{figure*}
This comparison shows that the third order Meijer-G approximants are excellent approximations to the exact Borel sum, and that fifth order Meijer-G approximants reproduce the exact Borel sum. Finally it is clear that the series is not Borel summable: using the Borel sum to estimate the value of $Z(g)$ results in a fictitious imaginary part with no clear physical interpretation.  The Borel sum fails to reproduce the analytic structure of $Z(g)$ on the complexified $g$-plane.


Cherman {\it et al.} \cite{Cherman2015} provide a very clear explanation of the appearence of the non-perturbative ambiguity in this problem, which echoes the discussion given above for the case of $\phi^4$ theory with degenerate vacua. In the present case, we have also an action $\frac{\sin(\phi)^2}{2g}$ with a perturbative saddle at $\phi=0$, with action $S_1=0$, as well as a non-perturbative saddle at $\phi=\pi/2$ where the action evaluates to $1/(4g)$ ($S_2=1/(4g)$). In order to properly evaluate $Z(g)$ we thus need to sum a transseries expansion of the form
\begin{equation}
Z(g,\sigma_0,\sigma_1)\sim \sigma_0 e^{S_0}\sum_{n=0}^\infty c_{n,0}g^n+\sigma_1 e^{S_1}\sum_{n=0}^\infty c_{n,1}g^n,
\end{equation}
where $c_{n,0}$ ( $c_{n,1}$) is the $n$-th order expansion coefficient around the perturbative (non-perturbative) saddle,  and $\sigma_0$ and $\sigma_1$ are the corresponding transseries parameters.  The transseries parameters have been found to be given by $\sigma_0=1$ and $\sigma_1=- \mathrm{i}$, for $\mathrm{Im}g>0$ and $\sigma_1=\mathrm{i}$, for $\mathrm{Im}g<0$.  The divergent series in the transseries expansion can both be summed exactly by Borel-Hypergeometric summation; the Borel-transformed series are both hypergeometric series
\begin{equation}
\sum_{n=0}^\infty \frac{c_{n,0}}{n !}\tau^n\sim \,_2F_1\left(\frac{1}{2},\frac{1}{2},1;2 \tau\right),
\end{equation}
and 
\begin{equation}
\sum_{n=0}^\infty \frac{c_{n,1}}{n !}\tau^n\sim \,_2F_1\left(\frac{1}{2},\frac{1}{2},1;-2 \tau\right),
\end{equation}
and thus belong to the space of functions that can be reconstructed by our technique. In particular, third order Meijer-G approximants cannot be exact, as the hypergeometric approximants in the Borel plane are of the form $_2F_1(1,h_1,h_2,h_3 \tau) $ rather than $_2F_1(1/2,h_1,h_2,h_3 \tau) $. But fifth order Meijer-G approximants can be exact since
\begin{equation}
_2F_1\left(\frac{1}{2},\frac{1}{2},1;2 \tau\right)=\,_3F_2\left(\frac{1}{2},\frac{1}{2},1,1,1;2 \tau\right),
\end{equation}
and therefore the Borel transformed series can be summed up exactly with fifth order Meijer-G approximants. In Figures ~\ref{Fig:fig6}(a)-(b) we compare Meijer-G (or Borel-Hypergeometric) approximants with the exact $Z(g)$.  In Fig~\ref{Fig:fig6}(a) we show the real part of $Z(g)$: the third order approximant already gives an excellent approximation to $\mathrm{Re}[Z(g)]$ while the fifth order approximant reproduces its exact value.    In Fig~\ref{Fig:fig6}(b) we see that both approximants have a non-perturbative ambiguity, which is again well approximated by the third order approximant. The non-perturbative ambiguity can be removed by the transseries summation by Meijer-G approximants, as shown in Figure~\ref{Fig:fig6}(c)-(d). The real part of $Z(g)$, shown in  Fig~\ref{Fig:fig6}(c), is unaffected but the non-perturbative ambiguity has been effectively removed.  Comparing  Fig.~\ref{Fig:fig6}(b)  with   Fig.~\ref{Fig:fig6}(d)  we see that third-order transseries summation dramatically reduced the size of the non-perturbative ambiguity.

For testing and pedagogical purposes we have chosen as a first application of our method the problem of approximating partition functions in 0-dimensional QFT. For those cases where the expansion is not Borel-summable we still use Meijer-G functions to approximate the Borel sums of the multiple divergent series that appear in a resurgent transseries. Such an approach can be described as the Borel-Hypergeometric-\'Ecalle summation method and for these examples delivers the exact answer at the fifth-order level, completely removing the non-perturbative ambiguity.  As a final application we discuss the computation of instanton corrections in the quantum mechanical double-well problem. 


\subsection{Summed-up one- and two-instanton contributions in the Double-Well Problem}
A more challenging case is provided by multi-instanton resummation in the quantum-mechanical double-well problem. This is a simple quantum-mechanical problem, but one which is not tractable by standard perturbation theory or by any resummation of perturbative data; a successful resummation requires the construction of a transseries (or multi-instanton expansion)~\cite{JentschuraMultinstanton} and here we use Meijer-G approximants in an attempt to sum such expansion.  

\begin{figure}
\center
\includegraphics[width=0.48\textwidth]{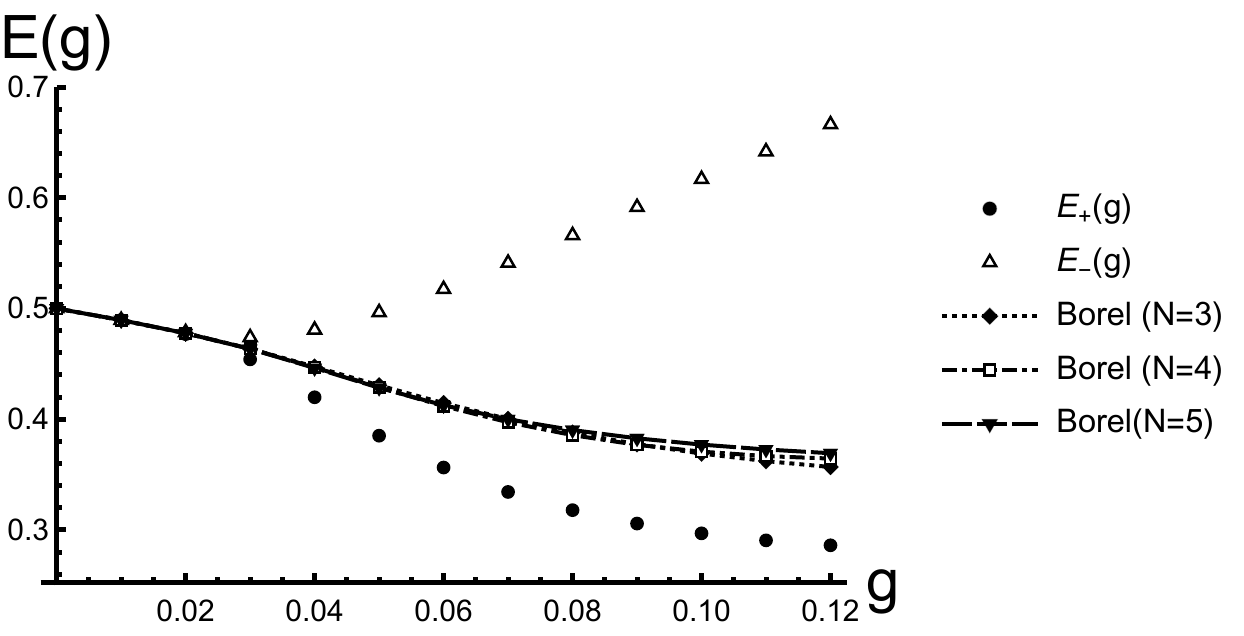}
\caption{The perturbation for the quantum-mechanical double-well problem is not Borel-summable. For $g= 0$ the ground state has two-fold degeneracy, corresponding to states with positive and negative parity ($E_+$ and $E_-$; filled circles and empty triangles, respectively). For $g\neq 0$ this degeneracy is broken and $E_+\neq E_-$. The energy gap that appears between the ground and first excited state is a truly non-perturbative effect, and cannot be accounted for by Borel summation (see legend). The Borel sums have been performed by Meijer-G approximants of orders three to five, and appear well converged for the range of values of $g$ shown.}
\label{Fig:fig7}
\end{figure}

The double-well potential is given by
\begin{equation}
V(x)=\frac{1}{2}x^2(1-\sqrt{g}x)^2.
\end{equation}
For $g=0$ we have a harmonic oscillator with energy $E_0=1/2$. However, as we increase $g$, the ground state splits into two different states with opposite parity. The eigenvalue of the lower energy state (first excited state) is denoted $E_+$ ($E_-$). The energy gap between these two states is a non-perturbative and not-Borel-summable quantity---it is exponentially small in the limit $g\rightarrow0$ and hence not detectable by standard perturbation theory.  This is shown in Fig.~\ref{Fig:fig7}, where $E_\pm$ are shown as  functions of $g$ together with Meijer-G approximations ($N=3,\,4,\,5$) to the Borel sum. Clearly the Meijer-G approximants succeed at rapidly converging the Borel sum, but the Borel sum is insufficient to describe the non-perturbative energy gap between $E_+$ and $E_-$. 
The multi-instanton expansion (or resurgent transeries) proposed by Zinn-Justin and Jenstchura \cite{JentschuraMultinstanton} can be used to estimate $E_{\pm}(g)$. Defining
\begin{eqnarray}
\xi&=&\frac{1}{\sqrt{\pi g}}e^{-\frac{1}{4g}},\\
\chi&=&\mathrm{ln}\left(-\frac{2}{g}\right),\\
\epsilon&=&\pm,
\end{eqnarray}
and neglecting multi-instanton contributions beyond the two-instanton order, $E_\epsilon(g)$ is approximated as
\begin{equation}
E_\epsilon(g)\approx E_0(g)+E^{(1)}_\epsilon(g)+E^{(2)}_\epsilon(g),
\label{eq:dw2instantonapp}
\end{equation}
where $ E_0(g)$ is the perturbation expansion, $E^{(1)}_\epsilon(g)$ is the one-instanton correction and $E^{(2)}_\epsilon(g)$ is the two-instanton correction. These are given by
\begin{widetext}
\begin{eqnarray}
E_0(g)&=&\frac{1}{2}-g-\frac{9}{2}g^2-\frac{89}{2}g^3-\frac{5013}{8}g^4+O(g^5),\label{eq:dw0}\\
E_\epsilon^{(1)}(g)&=&-\epsilon \xi(g) \left(1- \frac{71}{2}g-\frac{6299}{288}g^2-\frac{2691107}{10368}g^3-\frac{2125346615}{497664}g^4+O(g^5)\right),\label{eq:dw1}\\
E_\epsilon^{(2)}(g)&=&\xi^2(g) \chi(g)\left(1- \frac{53}{6}g-\frac{1277}{72}g^2-\frac{336437}{1296}g^3+O(g^4)\right)\nonumber \\
                            &+&\xi^2(g) \left(\gamma-\left( \frac{23}{2}+\frac{53}{6}\gamma\right)g-\left( -\frac{13}{12}+\frac{1277}{72}\gamma\right)g^2-\left( \frac{45941}{144}+\frac{336437}{1296}\gamma\right)g^3+O(g^4)\right)\label{eq:dw2},
\end{eqnarray}
\end{widetext}
where $\gamma$ is Euler's constant. We can see that this resurgent expansion consists of the standard perturbation expansion plus corrections; each of these corrections contains one or more divergent expansion in powers of $g$, which are multiplied by non-analytic terms like $\xi(g)$, $\xi^2(g)$ and $\xi(g)\chi(g)$. For instance the perturbation expansion in Eq.\eqref{eq:dw0} is a divergent series. Similarly for $E_\epsilon^{(1)}(g)$ in Eq.\eqref{eq:dw1}, the series that multiplies $-\epsilon \xi(g)$ is also divergent. Generalized {\it ad infinitum} a resurgent multi-instanton expansion can be described as a series in powers of $\xi(g)$ and $\chi(g)$ where the ``expansion coefficients'' are themselves divergent series that need to be resummed. The zeroth order of such expansion is the traditional perturbation theory. Higher order terms contain powers of the  form $\xi^n(g)\chi^m(g)$ multiplying a divergent series. 

In the case present case  contributions at odd instanton orders are parity dependent: they have opposite signs depending on the parity. In contrast, even-order  multi-instanton contributions are parity independent \cite{JentschuraMultinstanton}. Thus the one-instanton correction opens the gap between $E_+$ and $E_-$ as these states have opposite parity; the one-instanton correction lowers the energy of $+$ state, increasing the energy of $-$ state and therefore breaks the degeneracy observed at $g=0$. In contrast the two-instanton contribution tends to increase the energy of both $+$ and $-$ states.
\begin{figure*}
\center
\includegraphics[width=0.49\textwidth]{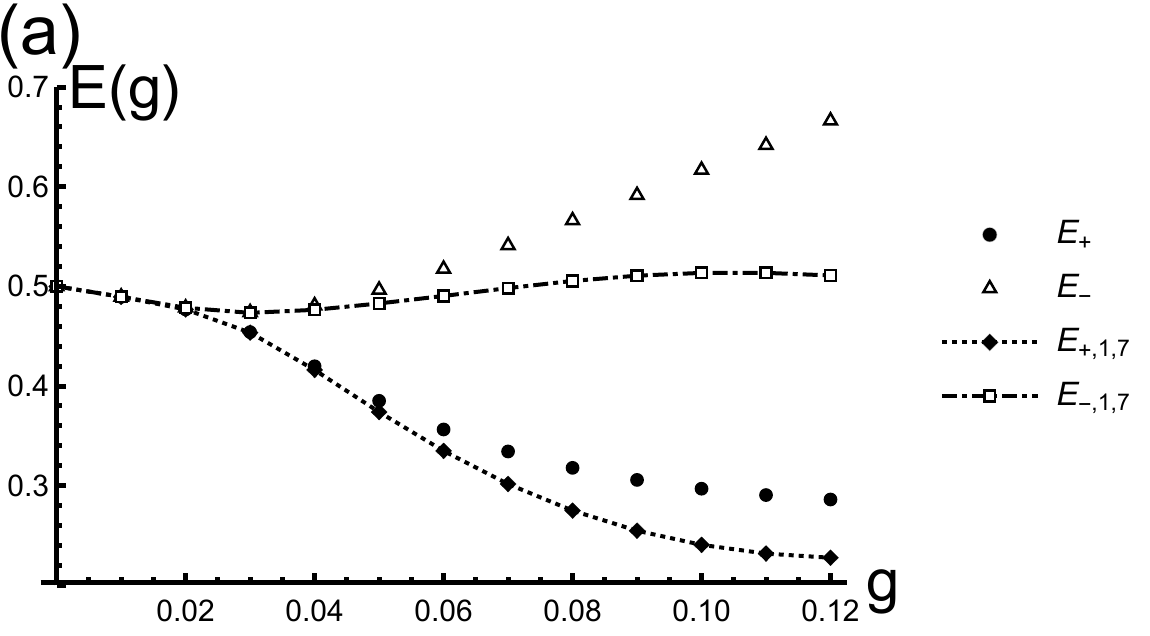}
\includegraphics[width=0.49\textwidth]{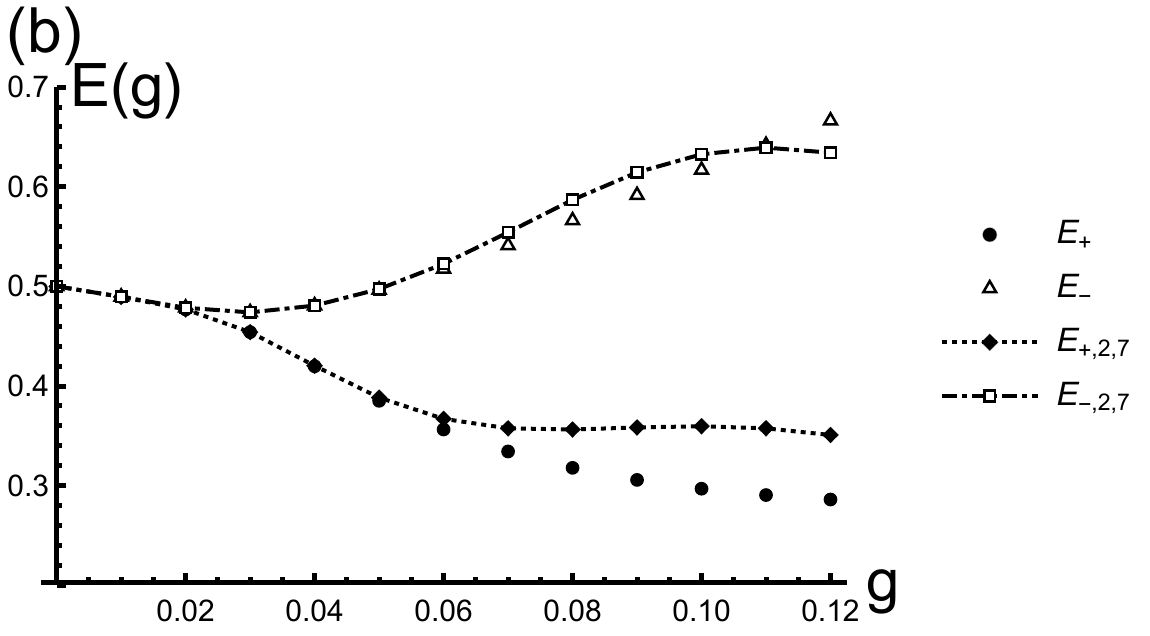}
\caption{Summation of the one- and two-instanton contributions to the ground and first excited state of the quantum mechanical double-well potential via seventh-order Meijer-G approximants. (a) Numerically obtained exact result [filled circles (ground state, $E_+$) and empty triangles (first excited state, $E_-$)], together with the one-instanton approximation as calculated by seventh-order Meijer-G approximants for the ground state [dotted line with filled diamonds ($E_{+,1,7}$)] and first excited state [dot-dashed line with empty squares ($E_{-,1,7}$)]; (b) The seventh-order Meijer-G two-instanton approximation compared with the exact result. While the Borel sum converges rapidly with increasing approximant order, the one- and two-instanton contributions converge at a slower rate. For $g=0.12$ one-instanton approximants of order $N\ge 5$ appear well converged. This is not the case for two-instanton Meijer-G approximations that are not fully converged for $N=$6--7.}
\label{Fig:fig8}
\end{figure*}

We will attempt to sum the resurgent transseries given by Eq.\eqref{eq:dw2instantonapp} by replacing each divergent series that contributes to  $E_0(g)$, $E^{(1)}_\epsilon(g)$ and $E^{(2)}_\epsilon(g)$ in Eqs. \eqref{eq:dw0}-\eqref{eq:dw2} by their respective Borel sums. The Borel sums will be calculated using Meijer-G approximants, including powers in $g$ of order seven or lower. Because the convergence of the approximants turned out to be slower than in the previous examples, here we calculate also the even-ordered approximants by starting the algortihm described above from the once-subtracted series.  For clarity we will show only results obtained by seventh-order Meijer-G approximants, describing briefly the apparent convergence of the approximants when necessary.  We denote the $N$-th order $n$-instanton approximations to $E_\epsilon$ as $E_{\epsilon,n,N}$; as mentioned above the $n$-instanton approximation are replaced by Meijer-G approximants of order $N\le 7$, i.e. we perform a Borel-Hypergeometric-\'Ecalle summation. For instance the third order one-instanton approximation is roughly given by
\begin{widetext}
\begin{equation}
E_{\pm,1,3}(g)\approx \frac{0.0288}{g}  \MeijerG*{3}{1}{2}{3}{0,-0.347}{-1.26,0,0}{-\frac{0.199}{g}}\mp  \frac{0.0222}{g} \xi(g)  \MeijerG*{3}{1}{2}{3}{0,-0.255}{-1.58,0,0}{-\frac{0.131}{g}}.
\end{equation}
\end{widetext}

In Fig.~\ref{Fig:fig7} we show how the coupling breaks the degeneracy between $+$ and $-$ states. Numerically exact values are compared with Meijer-G approximants to the Borel sum (zero-instanton) $E_{\pm,0,N}$ with $N=$3--5. The approximants appear to be well converged for the range of values of $g$ shown. The Borel sum does not account for the splitting between $+$ and $-$ states. In order to account for this non-perturbative effect we need multi-instanton corrections.  In Fig.~\ref{Fig:fig8} we compare the exact ground and first excited state energies with the one- and two-instanton seventh order approximation,---see Fig.~\ref{Fig:fig8}(a)---, $E_{\epsilon,1, 7}$ and $E_{\epsilon,2, 7}$. Multi-instanton corrections account for the non-perturbative gap opening. For the values of $g$ shown, the Meijer-G approximants appear to be well converged in the one-instanton case, while for the two instanton case they appear to be well converged  up to $g=0.08$ and are likely not inaccurate for all the values of $g$ shown.  From our forays at larger instanton orders we suspect that the convergence of the Meijer-G approximants gets slower with increasing instanton order. This is likely due to the fact that the Borel transform removes factorial growth, while the coeffieint growth in the divergent series that enter the multi-instanton expansion depend on instanton order, and they can grow superfactorially.  Does this trend continue at higher instanton orders? We leave the question of convergence of the summed-up multi-instanton expansion to future work.  In any case, this low order calculation shows that multi-instanton contributions open up the gap  and that the two-instanton approximation improves on  the one-instanton approximation. Once again Meijer-G approximants offer an economical, yet accurate, approach to  evaluate these corrections.

\begin{figure}
\center
\includegraphics[width=0.5\textwidth]{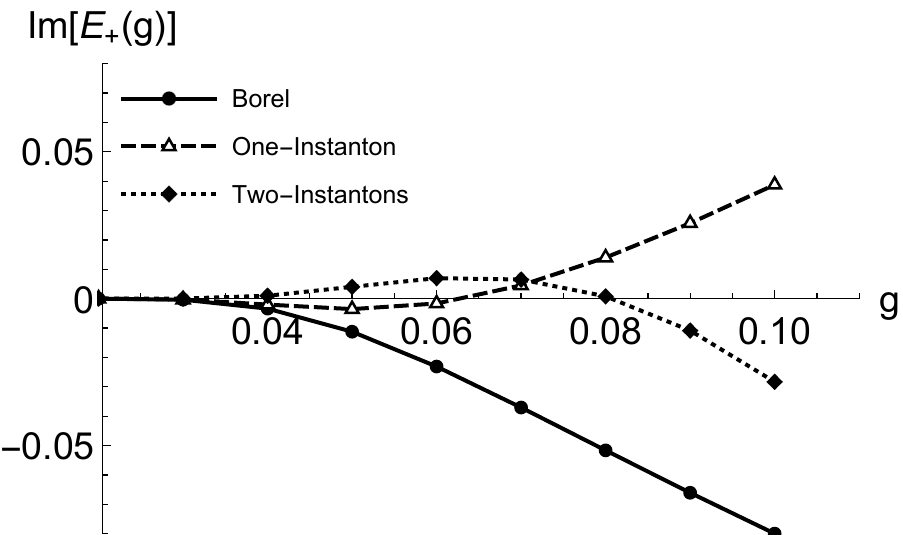}
\caption{Non-perturbative ambiguity as a function of coupling strength calculated for the quantum mechanical double-well problem. One- and two-instanton contributions reduce the size of the ambiguity but do not cancel entirely cancel it  in the interval $0.03<g<0.1$. }
\label{Fig:fig9}
\end{figure}

With the summed-up one- and two-instanton  approximations we can discuss the cancellation of the non-perturbative ambiguity as a function of $g$. We know that in the limit $g\rightarrow 0^+$ the non-perturbative ambiguity is cancelled by the two-instanton contribution, and left intact by the one-instanton contribution~\cite{JentschuraMultinstanton}.  However with a summed-up multi-instanton expansion  the ambiguity cancellation is far from trivial---way more complex than in the previous examples we have discussed. At the two-instanton order there are four different Borel sums, each of which introduces its own ambiguity; exact ambiguity cancellation at the two-instanton order only happens in a rather strict $g\rightarrow 0^+$ limit, and has been discussed in detail by Jentschura and Zinn-Justin~\cite{JentschuraMultinstanton}. Here we address the ambiguity cancellation at the two-instanton level, beyond the leading order cancellation present as $g\rightarrow 0^+$. In Fig.~\ref{Fig:fig9} we show  $\mathrm{Im}[E_+(g)]$ as a function of $g$.  Away from $g=0$ the ambiguity is reduced but not exactly cancelled. It is interesting to note that the one-instanton cancels part of the ambiguity away from $g\rightarrow 0$; away from $g\rightarrow 0^+$ the two-instanton contribution does not result in a marked improvement relative to the one-instanton result.   
 
 In line with the results of previous sections, the Meijer-G approximants  are then able to provide good approximations to the Borel sum and one-instanton correction, apparently converging quickly with increasing order. We have observed that the convergence of the approximants becomes slower with increasing instanton order. The cancellation of the non-perturbative ambiguity has been studied for $g>0$ and appears fragile: each of the Borel sums performed by Meijer-G summation at each instanton order  has its own ambiguity and the cancellation is not as neat as in the previous examples, where the transseries contained only two terms.  Away from $g=0$ the one-instanton contribution reduces the size of the ambiguity, but the two-instanton contribution does not yield a marked improvement relative to one-instanton calculations. In contrast as $g\rightarrow 0^+$ the two-instanton contribution cancels exactly the non-perturbative ambiguity, while the one-instanton contribution misses this cancellation entirely.

\section{Discussion: is Borel-Pad\'e  obsolete?}\label{Sec:IV}
We have put forward a resummation approach that apparently surpasses the commonly used Borel-Pad\'e method in accuracy and ease of use. So a natural question to ask is whether we have actually put the final nail in the coffin of Borel-Pad\'e approaches. The short answer to this question is...not quite. While there are very clear advantages in adopting Meijer-G approximants there are also some disadvantages that still need to be alleviated. A few of these advantages are:
\begin{enumerate}
\item {\bf Meijer-G approximants are easily parameterized.} The hypergeometric ansatz allows for a swift parameterization of Meijer-G approximants. By approximating the ratios between consecutive Borel-transformed coefficients by a rational function, one parameterizes a hypergeometric function in the Borel plane. The Laplace transform of any hypergeometric function is known analytically in terms of Meijer-G functions.  The algorithm given in Section is then an easy recipe to transform the coefficients of a divergent perturbation expansion into tables of Meijer-G approximants. One only needs to find the rational functions $r_N(n)$ that approximate ratios between consecutive coefficients in the Borel plane and from them the hypergeometric vectors;  such program involves only the solution of a set of linear equations in the former case; and the calculation of polynomial roots in the latter case.
\item {\bf Meijer-G approximants are more accurate than Borel-Pad\'e approximants.} The examples given here, plus a number of other problems we have considered over the course of this work, lead us to be believe that this is clearly the case when the convergence-limiting singularity in the Borel plane is a branch cut. Here we have shown how the Meijer-G approximants actually converge to the exact Borel sum in various examples from QFT. 
\item {\bf Meijer-G functions also approximate generalized Borel sums.} Why should we be limited to Borel transforms of the form $b_n=z_n/n!$? Or Mittag-Leffler transforms of the form $b_n=z_n/\Gamma(a+b n)$? Work in progress shows that more general transforms can be defined that are able to remove, in essence, arbitrary asymptotic coefficient growth and that the resulting approximants can also be represented in terms of Meijer-G functions. Generalizations of the Borel sum can easily be constructed by means of Meijer-G functions.
\end{enumerate}
But there are disadvantages too:
\begin{enumerate}
\item {\bf Meijer-G approximants are new} and therefore their convergence properties are unknown. In contrast the properties of Pad\'e approximants are well understood.
\item  {\bf Meijer-G approximants can be difficult to evaluate numerically,} particularly at large orders. Our approach involves the numerical evaluation of Meijer-G functions, using black boxes which may have not been fully tested at very large orders  (who ever needed $_{11}F_{10}$?). To obtain accurate Meijer-G approximations one typically needs high-accuracy input data, which may be not available.
\item  {\bf Pad\'e approximants are better for poles.} Meijer-G approximants are very well suited for problems where the Borel plane has a branch cut. But this is not always the case. In some cases the perturbation expansion, or its Borel-transformed counterpart, may have its convergence limited by a pole. Physical examples of such behaviour include perturbative spectral functions in Green's function theory~\cite{Pavlyukh2017} and the beta function of supersymmetric Yang-Mills theories~\cite{Novikov1983}. In such cases Pad\'e or Borel-Pad\'e approximants are likely a better summation method than hypergeometric or Borel-Hypergeometric summation. Our approach therefore complements, but does not replace, Pad\'e and Borel-Pad\'e summation. 
\end{enumerate}
Our work then calls for further investigations on the convergence properties of Meijer-G approximants, and extensive stress-testing of numerical black boxes for Meijer-G function evaluation. A more fundamental problem is the development of generalized Borel summation methods and parameterizing the corresponding  Meijer-G  approximants.   Finally it will be interesting to see how such summation approach, based on high-end special functions, would fare in a real-world scenario where only very few coefficients of limited accuracy are available.  


\section{Conclusions}\label{Sec:V}
To conclude, we put forward a simple algorithm that enables a fast and accurate low-order Borel summation. The algorithm is a generalized Borel-Hypergeometric approach, where the hypergeometric ansatz is used to transform the coefficients of a divergent series into a table of Hypergeometric approximants in the Borel plane, and hence into a table of Meijer-G approximants to the Borel sum of the series. We succesfully applied this technique to the summation of divergent series and resurgent transseries by Meijer-G approximants.  We have considered as examples various partition functions in 0-dimensional QFT; in these cases the Meijer-G approximants converge to the exact answer at order $N=5$. We have used these approximants to sum transseries (Borel-Hypergeometric-\'Ecalle sum), completely removing the non-perturbative ambiguity in the Borel sums. The summation of the multi-instanton expansion for the quantum mechanical double-well problem provided a more challenging example, where the convergence of the Meijer-G approximants was found to be slower. Nevertheless the Meijer-G approximants put forward here were able to yield accurate low order approximations to the Borel sum and the one- and two-instanton corrections.

\begin{acknowledgments}
H.~M. and T.~G.~P. acknowledge financial support from the CNG center under the Danish National Research Foundation, project DNRF103. T.~G.~P. was also supported by the QUSCOPE center sponsored by the Villum Foundation. H.~M. and B.~K~.N.  were supported by NSF Grant No. CHE 1566074. 
\end{acknowledgments}

\end{document}